\documentclass[preprint,12pt]{elsarticle}

\usepackage{amsmath,amssymb}
\usepackage{graphicx}
\usepackage{hyperref}
\usepackage{siunitx}
\usepackage{booktabs}
\usepackage{color}

\journal{Next Research}

\begin{document}
	
	\begin{frontmatter}
		
		\title{Three-dimensional photon transport in spinodal photocatalytic aerogels:\\
			how bicontinuous morphology controls kinetic rate constants}
		
		\author[bordeaux]{Renaud A.L. Vall\'ee\corref{cor1}}
		\cortext[cor1]{Corresponding author.}
		\ead{renaud.vallee@crpp.cnrs.fr}
		\address[bordeaux]{Univ.\ Bordeaux, CNRS, CRPP, UMR5031, 33600 Pessac, France}
		
		\begin{abstract}
			Porous monolithic photocatalysts based on anatase TiO$_2$ dispersed in a silica
			aerogel matrix are promising platforms for indoor air purification and VOC
			photo-degradation. Their bicontinuous spinodal architecture simultaneously
			provides high specific surface area, structural integrity, and intense multiple
			scattering of UV photons. Extracting intrinsic kinetic descriptors from measured
			apparent rate constants, however, requires an optical model for the photon
			fluence inside the material, and current practice replaces the complex
			three-dimensional (3D) pore network by a homogeneous one-dimensional (1D) slab
			with effective optical coefficients---an approximation whose error has never been
			quantified for spinodal geometries. Here we combine realistic 3D spinodal pore
			masks generated by Cahn--Hilliard spectral integration with full 3D
			GPU-accelerated Monte Carlo (MC) photon-migration simulations. We introduce a
			solid-phase-weighted fluence estimator $\langle\Phi\rangle_{\text{MC,3D}}$ that
			accounts for the spatial distribution of catalytic sites and compare it with the
			conventional volume average $\langle\Phi\rangle_{\text{MC,vol}}$ and with the
			diffusion-approximation (DA) prediction $\langle\Phi\rangle_{\text{DA}}$. The
			ratio $\langle\Phi\rangle_{\text{MC,3D}}/\langle\Phi\rangle_{\text{MC,vol}}$
			equals 1.515 at porosity $\phi = 0.70$ and 1.774 at $\phi = 0.90$: the catalytic
			solid phase is preferentially illuminated. Depth-resolved analysis identifies the
			mechanism as an optical \emph{phase contrast}. Because the pores neither absorb
			nor scatter, photons cross them quasi-ballistically and the void phase sustains
			only $r = 0.514$ of the fluence found in the scattering solid, so that the
			conventional volume average is diluted by dark pore voxels. The ratio $r$ is found empirically to be the same to within 0.2\% at the two
			simulated porosities; the concentration ratio then follows from the exact
			relation $1/[\phi r + (1-\phi)]$.
			The kinetic descriptor $k_{\text{surf}}\cdot K_{\text{ads}}$ extracted from the
			measured apparent rate constant is underestimated by 20\% along the DA route and
			overestimated by 51.5\% along the volume-averaged route. Homogeneous-slab MC
			simulations at matched optical thickness give a fluence-weighted RMS discrepancy
			of 8.5\%, compared with 54.6\% for the explicit spinodal geometry. The much
			larger discrepancy shows that the homogeneous DA benchmark does not adequately
			represent transport in the simulated bicontinuous structure. These figures
			compare two empirical metrics; they are not an additive decomposition of the
			spinodal MC--DA error. These results provide a quantitative correction for
			kinetic extraction in bicontinuous photocatalysts and quantitative design
			considerations linking porosity to photon utilisation efficiency.
		\end{abstract}
		
		
	\end{frontmatter}
	
	\begin{highlights}
		\item Solid-phase fluence exceeds the volume average by 51.5\% ($\phi = 0.70$) and 77.4\% ($\phi = 0.90$)
		\item Mechanism is an optical phase contrast, $\langle\Phi\rangle_{\text{pore}}/\langle\Phi\rangle_{\text{solid}} = 0.514$, unchanged over the two simulated porosities
		\item \item Volume averaging overestimates $k_{\text{surf}}\cdot K_{\text{ads}}$
		by 51.5\% at $\phi = 0.70$; the DA underestimates it by 20\%
		\item At matched optical thickness, $\mathrm{RMS}_w$ is 8.5\% for a homogeneous
		EMA slab and 54.6\% for the spinodal geometry
		\item At $\phi = 0.90$, 14\% of escaping photons travel further inside the void network than the slab is thick
	\end{highlights}
	
	\section{Introduction}
	\label{sec:intro}
	
	Semiconductor photocatalysis using TiO$_2$ has emerged as one of the most
	versatile approaches for environmental remediation, from water purification to
	indoor air treatment and solar-driven synthesis~\cite{Hoffmann1995}. Among the
	many reactor configurations explored, porous monolithic photocatalysts based on
	anatase TiO$_2$ nanoparticles (NPs) dispersed in a silica aerogel matrix have
	attracted growing attention~\cite{Vardon2025,Layan2025}. Their macroporous,
	bicontinuous architecture offers simultaneously a high specific surface area
	accessible to gaseous reactants, structural integrity, and intense multiple
	scattering of UV photons---three conditions essential for efficient heterogeneous
	photocatalysis~\cite{Hoffmann1995,Ollis2018}.
	
	Understanding and controlling photon transport inside these materials is a
	prerequisite for rational design. The apparent rate constant $k_{\text{app}}$
	measured in a degradation experiment integrates three coupled phenomena:
	(i)~photon transport and absorption within the porous volume, (ii)~generation and
	surface migration of photo-excited charge carriers, and
	(iii)~Langmuir--Hinshelwood surface reaction kinetics~\cite{Ollis2018,Bloh2019}.
	To extract the intrinsic surface reactivity from $k_{\text{app}}$, one must
	therefore know the local photon fluence $\Phi(\mathbf{r})$ inside the monolith.
	This fluence is commonly estimated using a one-dimensional (1D) diffusion
	approximation (DA) with Maxwell--Garnett effective-medium optical
	coefficients~\cite{Garnett1904,Ishimaru1978}. Our companion
	paper~\cite{Vallee2026} derived compact closed-form expressions for the intrinsic
	volumetric rate constant $k_{V,\text{mono}}$ within this framework and validated
	them against 1D Monte Carlo (MC) simulations.
	
	What the 1D benchmark cannot assess is the role of the actual three-dimensional
	pore morphology. Real Si(HIPE) aerogels~\cite{Vardon2025} display a bicontinuous
	spinodal microstructure with interconnected solid and void phases, characterised
	by a broad pore-size distribution and topological features---percolation,
	saddle-point interfaces, and correlated void channels---that have no counterpart
	in a 1D slab model. Several studies have highlighted that microstructure geometry
	affects both the effective optical coefficients~\cite{Vynck2023} and the photon
	path-length distribution in porous media~\cite{Leyre2020,Leyre2022}, while
	topology-optimised 3D reactors have demonstrated that pore connectivity controls
	photon mean free path and optical thickness independently~\cite{Brunser2023}.
	A direct quantification of the morphological effect on photocatalytic rate
	constants, comparing a geometrically faithful 3D simulation with the analytical
	1D prediction, has however not been reported.
	
	Breaking the one-dimensional homogenisation assumption is part of a broader,
	cross-disciplinary shift in optical modelling and analysis, in which the
	three-dimensional spatial distribution of a component---rather than its mean
	content---is recognised as the controlling micro-indicator of function.
	In biomedical optics, voxel-based three-dimensional Monte Carlo modelling of
	photon migration through anatomically realistic, optically heterogeneous tissue
	has long been used to quantify spatial fluence distributions that homogeneous
	or layered approximations cannot capture~\cite{Boas2002,Fang2009}; the same
	class of solvers has recently been applied to transcranial photobiomodulation,
	where a voxelized anatomical head model quantifies photon penetration into
	the brain parenchyma~\cite{Zhang2026PBM}. Parallel developments in optical imaging show
	that spatial statistics themselves can be the diagnostic observable:
	AI-assisted multimodal nonlinear optical imaging has identified
	\emph{protein spatial homogenization} as a hallmark of impaired bone quality in
	type-2 diabetes~\cite{Zhang2026}, a finding that mean protein content alone
	does not encode---conceptually the same distinction drawn here between a
	solid-phase-weighted fluence and a conventional volume average. Conversely,
	engineered spatial heterogeneity underpins light management in correlated
	disordered photonic materials~\cite{Vynck2023}. In the same
	spirit, the present work shows that, at fixed composition and porosity, it is the
	3D spatial arrangement of the solid skeleton that controls both the optical
	response and the kinetic descriptors extracted from photocatalytic measurements.
	
	This gap matters because the field of photocatalysis is moving toward data-driven
	and rational design strategies~\cite{Liu2022,Bloh2019}, which require accurate
	kinetic descriptors. If the optical model used to extract these descriptors
	introduces a systematic error of several tens of percent, the resulting
	structure--activity relationships will be unreliable.
	
	The present paper addresses this with five specific objectives: (i)~generate
	realistic 3D spinodal aerogel geometries by numerical integration of the
	Cahn--Hilliard equation~\cite{Cahn1958,Cahn1959,Provatas2010}; (ii)~run full 3D
	GPU Monte Carlo photon-migration simulations~\cite{Fang2009,Yu2018} with an
	exactly closing photon budget; (iii)~introduce a solid-phase-weighted fluence
	estimator that accounts for the spatial distribution of catalytic sites;
	(iv)~compare fluence profiles and intrinsic rate constants across three
	estimators (MC~3D solid-weighted, MC~vol volume-averaged, DA collimated
	effective-medium), identifying where and why the approaches diverge; and
	(v)~establish, from phase-resolved fluence profiles and per-photon path-length
	statistics, the physical mechanism responsible for the divergence.
	
	The paper is organised as follows. Section~\ref{sec:theory} presents the complete
	kinetic chain from photon flux to photocatalytic rate constant, the spinodal
	geometry generation, the optical property calculation, the discrepancy metric and
	the simulation workflow. Section~\ref{sec:results} presents the pore morphology,
	the photon energy balance, the fluence profiles and 2D map, the quantitative
	signatures of the transport mechanism, the kinetic rate constants and the
	uncertainty budget. Section~\ref{sec:discussion} interprets the mechanism,
	confronts the predicted optical response with the measurements available on the
	companion experimental system, discusses DA validity and derives reactor design
	implications.
	Section~\ref{sec:conclusions} summarises the main conclusions. Two appendices
	give the complete numerical parameters of the geometry generator
	(\ref{app:CH}) and of the transport solver (\ref{app:MC}).
	
	\section{Theory and methods}
	\label{sec:theory}
	
	\subsection{From photon flux to photocatalytic rate: the complete kinetic chain}
	\label{sec:kinetic_chain}
	
	We first lay out the complete chain connecting the incident irradiance to the
	measurable apparent rate constant, with full dimensional analysis at each step.
	
	\subsubsection{Incident photon flux}
	
	At the entrance face of the monolith, the \emph{incident} photon flux is
	\begin{equation}
		S_0 = E_{\text{in}} = \frac{I_0}{h\nu},
		\qquad [S_0] = \text{photon}\;\text{m}^{-2}\;\text{s}^{-1},
		\label{eq:S0}
	\end{equation}
	where $E_{\text{in}}$ is the incident photon irradiance and $I_0$ the incident
	power density. For a collimated UV source at $\lambda = 365$~nm with
	$I_0 = 45$~W\,m$^{-2}$, one obtains
	$S_0 = 8.27 \times 10^{19}$~photon\,m$^{-2}$\,s$^{-1}$.
	
	The external Fresnel reflectance is deliberately \emph{not} folded into $S_0$,
	because it is handled natively by the Monte Carlo solver: the one-voxel air pad
	places the air/solid interface inside the computational volume, and the internal
	Fresnel treatment (\texttt{isreflect = 1}, Section~\ref{sec:MC} and
	Appendix~\ref{app:MC}) reduces the weight of the packets reflected there. Packets
	are therefore launched with unit weight, and $S_0$ of Eq.~\eqref{eq:S0} is the
	flux \emph{prior} to surface reflection---which is also the quantity used to
	convert the solver output into absolute fluence in Eq.~\eqref{eq:Phi_norm}.
	Because the entrance face is a fraction $\phi$ of air and $(1-\phi)$ of solid,
	the resulting entry loss is an \emph{area-weighted estimate}
	$(1-\phi)R_{\mathrm{F}}(n_{\text{eff}})$---about 2\% at $\phi = 0.70$ and 0.7\%
	at $\phi = 0.90$ for $n_{\text{eff}} = 1.746$ at normal incidence---not an exact
	geometrical consequence of bulk porosity, and it is already contained in the
	measured reflectance $R$ of Table~\ref{tab:energy}. The DA uses the same
	incident-flux normalisation. Its ballistic source does not include an explicit
	entry-transmission factor, and the adopted extrapolation length
	$z_b = (2/3)\ell^*_{\text{eff}}$ is the index-matched value, with no
	internal-reflection factor from Haskell et al. This is not a partial
	implementation of the same Fresnel treatment used in the Monte Carlo. The
	difference in source and boundary treatment may contribute to the near-surface
	MC--DA deviation and to the residual discrepancy of the homogeneous-slab tests
	(Section~\ref{sec:homogeneous}).
	
	\subsubsection{Local photon fluence and volumetric absorption}
	
	Inside the porous monolith, the local photon fluence $\Phi(\mathbf{r})$---defined
	as the angularly integrated photon flux---obeys the radiative transfer equation.
	In the diffusion regime ($\mu_s' \gg \mu_a$) it satisfies~\cite{Vallee2026}
	\begin{equation}
		-D\,\nabla^2\Phi(\mathbf{r}) + \mu_a\,\Phi(\mathbf{r}) = S(\mathbf{r}),
		\label{eq:diffusion_general}
	\end{equation}
	where $D = 1/[3(\mu_a + \mu_s')]$ is the optical diffusion coefficient, $\mu_a$
	and $\mu_s'$ the absorption and reduced scattering coefficients,
	$\ell^* = 1/\mu_s'$ the transport mean free path, and $S(\mathbf{r})$ a
	\emph{volumetric} source term. Note the dimensional distinction: $S_0$ in
	Eq.~\eqref{eq:S0} is a \emph{surface} flux
	($\text{m}^{-2}\,\text{s}^{-1}$), whereas $S(\mathbf{r})$ is a \emph{volumetric}
	source ($\text{m}^{-3}\,\text{s}^{-1}$); they differ by one length dimension and
	cannot be compared directly.
	
	The local volumetric absorption rate is
	\begin{equation}
		q_a(\mathbf{r}) = \mu_a\,\Phi(\mathbf{r}),
		\qquad [q_a] = \text{photon}\;\text{m}^{-3}\;\text{s}^{-1}.
		\label{eq:qa}
	\end{equation}
	
	\subsubsection{Carrier generation and surface reaction}
	
	Not all absorbed photons produce a useful electron--hole pair. The internal
	quantum yield $\phi_{\text{int}}$ accounts for this:
	\begin{equation}
		G(\mathbf{r}) = \phi_{\text{int}}\,\mu_a\,\Phi(\mathbf{r}),
		\qquad [G] = \text{carrier}\;\text{m}^{-3}\;\text{s}^{-1}.
		\label{eq:G}
	\end{equation}
	Typically $\phi_{\text{int}} \in [0.1,\,0.3]$ for TiO$_2$ anatase under
	UV~\cite{Hoffmann1995}.
	
	The photocatalytic reaction occurs at the surface of the TiO$_2$ NPs. Postulating
	a site-limited mechanism,
	\begin{equation}
		r_A(\mathbf{r}) = k_{\text{surf}}\,G(\mathbf{r}),
		\qquad [r_A] = \text{molecule}\;\text{m}^{-2}\;\text{s}^{-1},
		\label{eq:rA}
	\end{equation}
	dimensional consistency imposes $[k_{\text{surf}}] = \text{m}$. This is unusual
	for chemists accustomed to surface rate constants in m\,s$^{-1}$; the difference
	arises because $G(\mathbf{r})$ is a volumetric generation rate, not a surface
	concentration. Physically, $k_{\text{surf}}$ is an \emph{effective capture
		length}: the typical thickness of volume from which a photo-generated carrier can
	reach the surface and react. For NPs of radius $a_{\text{np}} \approx 7.25$~nm,
	the carrier diffusion length in nanocrystalline anatase ($L_D \sim 1$--30~nm) is
	comparable to $a_{\text{np}}$, so that every carrier generated within a NP can
	reach its surface. One can then write~\cite{Hoffmann1995}
	\begin{equation}
		k_{\text{surf}} \sim a_{\text{np}} \cdot p_{\text{react}} \cdot \chi_{\text{access}},
		\label{eq:ksurf_micro}
	\end{equation}
	where $p_{\text{react}}$ is the probability of reaction at each carrier--surface
	contact and $\chi_{\text{access}}$ is the fraction of the NP surface exposed to
	the pollutant-bearing fluid. The value
	$k_{\text{surf}} = a_{\text{np}} = 7.25 \times 10^{-9}$~m used below corresponds
	to the upper bound $p_{\text{react}} \cdot \chi_{\text{access}} = 1$. In practice
	a significant fraction of NPs may be embedded in the silica matrix and
	inaccessible, so that $p_{\text{react}} \cdot \chi_{\text{access}} \sim 10^{-2}$
	is physically plausible. Importantly, $k_{\text{surf}}$ thus becomes an indirect
	probe of NP accessibility within the aerogel---exactly the parameter that
	synthesis seeks to optimise.
	
	\subsubsection{Accessible catalytic surface}
	
	The accessible catalytic surface per unit volume is
	\begin{equation}
		S_{\text{acc}} = \frac{S_{\text{cat}}}{V_{\text{dom}}}
		= f_{\text{NP}}\,(1-\phi)\,\frac{3}{a_{\text{np}}},
		\qquad [S_{\text{acc}}] = \text{m}^{-1},
		\label{eq:Sacc}
	\end{equation}
	where $\phi$ is the aerogel porosity. This quantity is directly related to the
	BET specific surface area $S_{\text{BET}}$ (in m$^2$\,g$^{-1}$) via
	$S_{\text{acc}} = S_{\text{BET}} \times \rho_{\text{app}}$, where
	$\rho_{\text{app}}$ is the apparent density of the monolith. Because both the
	catalytic reaction and the photon absorption take place exclusively in the solid
	phase, it is convenient to introduce the same quantity referred to the
	\emph{solid} volume rather than to the total domain volume,
	\begin{equation}
		S_{\text{acc}}^{\text{solid}} \equiv \frac{S_{\text{cat}}}{V_{\text{solid}}}
		= \frac{S_{\text{acc}}}{1-\phi}
		= f_{\text{NP}}\,\frac{3}{a_{\text{np}}}
		= 1.034 \times 10^{8}~\text{m}^{-1},
		\label{eq:Sacc_solid}
	\end{equation}
	which is independent of porosity. Both densities are in m$^{-1}$, so the choice between them is not constrained by
	dimensional analysis; the condition that the solid fraction appear once in the
	product converting fluence into a mean rate per unit total volume is set out
	explicitly after Eq.~\eqref{eq:ksKads_extract}.
	
	\subsubsection{Volumetric reaction rate and intrinsic rate constant}
	
	The volumetric reaction rate, evaluated locally inside the solid phase, is
	\begin{equation}
		r_V(\mathbf{r}) = S_{\text{acc}}^{\text{solid}}\,k_{\text{surf}}\,
		\phi_{\text{int}}\,\mu_a\,\Phi(\mathbf{r}),
		\qquad [r_V] = \text{molecule}\;\text{m}^{-3}\;\text{s}^{-1},
		\label{eq:rV}
	\end{equation}
	where $\mu_a$ is the shell absorption coefficient of Table~\ref{tab:optics} and
	$\Phi(\mathbf{r})$ the local fluence in the solid: surface density, absorption
	coefficient and fluence are all referred to the solid volume, so that $r_V$ is a
	reaction rate per unit solid volume.
	Because reaction and absorption occur only in the solid phase,
	$V_{\text{solid}} = (1-\phi)V$, converting this solid-referenced local rate into
	a mean rate per unit \emph{domain} (total) volume---needed both for the
	intrinsic rate constant introduced below and for the kinetic extraction of
	Section~\ref{sec:kinetic_extraction}---requires an explicit volume average:
	\begin{align}
		r_{V,\text{mean}}
		&= \frac{1}{V}\int_{V_{\text{solid}}}
		S_{\text{acc}}^{\text{solid}}\,k_{\text{surf}}\,\phi_{\text{int}}\,
		\mu_a\,\Phi(\mathbf{r})\,\mathrm{d}V \notag \\
		&= S_{\text{acc}}^{\text{solid}}\,k_{\text{surf}}\,\phi_{\text{int}}\,\mu_a\,
		\frac{V_{\text{solid}}}{V}\,
		\frac{1}{V_{\text{solid}}}\int_{V_{\text{solid}}}\Phi(\mathbf{r})\,\mathrm{d}V
		\notag \\
		&= S_{\text{acc}}^{\text{solid}}\,k_{\text{surf}}\,\phi_{\text{int}}\,
		\underbrace{\mu_a\,(1-\phi)}_{\displaystyle \mu_{a,\text{eff}}}\,
		\langle\Phi\rangle
		\;\equiv\; S_{\text{acc}}\,k_{\text{surf}}\,\phi_{\text{int}}\,\mu_a\,
		\langle\Phi\rangle .
		\label{eq:rVmean_norm}
	\end{align}
	A single factor $(1-\phi)$ appears, carried by $\mu_{a,\text{eff}}$.
	Consistency does not require the surface density and the fluence average to
	share a reference volume; it requires that the solid fraction appear once.
	Using $S_{\text{acc}}$ (Eq.~\eqref{eq:Sacc}, which already carries $(1-\phi)$)
	together with $\mu_{a,\text{eff}}$ would apply that factor twice and
	underestimate $r_{V,\text{mean}}$ by $(1-\phi)$.
	
	The companion paper~\cite{Vallee2026} introduced an absorption-normalised
	intrinsic monochromatic volumetric rate constant of dimension length, obtained
	from $r_{V,\text{mean}}$ by removing the dependence on the incident flux $S_0$
	and on the material's absorption strength $\mu_a$ at the working wavelength:
	\begin{equation}
		\boxed{
			k_{V,\text{mono}} \equiv \frac{L}{\mu_a\,S_0}\,r_{V,\text{mean}}
			= k_{\text{surf}}\,S_{\text{acc}}\,\phi_{\text{int}}\,
			\frac{\langle\Phi\rangle L}{S_0}
			= k_{\text{surf}}\,S_{\text{acc}}\,\phi_{\text{int}}\,
			L_{\text{opt}},
		}
		\label{eq:kV}
	\end{equation}
	where $L_{\text{opt}} = \int_0^L \Phi(z)\,\mathrm{d}z / S_0 =
	\langle\Phi\rangle L/S_0$ is the effective optical length. Thus
	$k_{V,\text{mono}}$ is not itself a reaction rate: it is $r_{V,\text{mean}}$
	expressed as a photon-flux- and absorption-independent length, useful for
	comparing materials of different $\mu_a$; the 3D kinetic conclusions below use
	the dimensional product $k_{\text{surf}}\cdot K_{\text{ads}}$
	(Eq.~\eqref{eq:ksKads}), extracted directly from $r_{V,\text{mean}}$ without
	this extra normalisation. Dimensional analysis confirms
	$[k_{V,\text{mono}}] = \text{m}$.
	
	Substituting the DA solution for $\Phi(z)$ (valid when $\mu_a \ll \mu_s'$) yields
	the compact form of the companion paper,
	\begin{equation}
		k_{V,\text{mono}}
		= k_{\text{surf}}\,S_{\text{acc}}\,\phi_{\text{int}}\,
		\frac{\ell^*}{\displaystyle 1 + \frac{L}{2\ell^*} + \frac{z_b}{\ell^*}},
		\label{eq:kV_compact}
	\end{equation}
	where $z_b = \frac{2}{3}\ell^*$ is the extrapolation length and $\ell^*$ is the
	EMA transport mean free path. The denominator
	$1 + L/(2\ell^*) + z_b/\ell^*$ is the normalised effective optical thickness: the
	thicker the slab relative to $\ell^*$, the more photons are diluted and the
	smaller $k_{V,\text{mono}}$. The numerator shows that increasing the accessible
	surface and lengthening the transport mean free path both improve efficiency.
	
	\subsubsection{Extraction of $k_{\text{surf}} \cdot K_{\text{ads}}$}
	
	Under dilute Langmuir--Hinshelwood conditions ($K_{\text{ads}}\,C \ll 1$), the
	kinetics reduce to pseudo-first order~\cite{Ollis2018}:
	\begin{equation}
		\frac{\mathrm{d}C}{\mathrm{d}t} = -k_{\text{app}}\,C,
		\qquad
		k_{\text{app}} = r_{V,\text{mean}}\,K_{\text{ads}},
		\label{eq:kapp}
	\end{equation}
	where $r_{V,\text{mean}}$ is the domain-mean volumetric rate of
	Eq.~\eqref{eq:rVmean_norm}, averaged over the \emph{total} monolith volume.
	Without independent knowledge of $K_{\text{ads}}$, the experiment gives access
	only to the product~\cite{Bloh2019}
	\begin{equation}
		\boxed{
			k_{\text{surf}} \cdot K_{\text{ads}}
			= \frac{k_{\text{app,exp}}}
			{S_{\text{acc}}^{\text{solid}}\,\phi_{\text{int}}\,\mu_{a,\text{eff}}\,
				\langle\Phi\rangle},
			\qquad [k_{\text{surf}} \cdot K_{\text{ads}}]
			= \text{m}^4\;\text{photon}^{-1}.
		}
		\label{eq:ksKads}
	\end{equation}
	This is the intrinsic measurable quantity that allows comparison across materials
	of different morphologies. Crucially, the value extracted depends on \emph{which}
	$\langle\Phi\rangle$ estimator is used---and this is precisely where the 3D
	morphology enters.
	
	\subsection{Spinodal geometry generation via Cahn--Hilliard}
	\label{sec:cahn_hilliard}
	
	The three-dimensional pore mask is generated by numerical integration of the
	Cahn--Hilliard equation~\cite{Cahn1958,Cahn1959}, which describes spinodal
	decomposition---the mechanism by which a homogeneous mixture spontaneously
	separates into two interpenetrating phases without nucleation barrier. This
	equation governs the formation of bicontinuous morphologies in a wide range of
	materials, from polymer blends to metallic alloys~\cite{Provatas2010,Geslin2015}
	and, relevant here, sol--gel-derived silica aerogels~\cite{Vardon2025}.
	
	The local solid volume fraction $\varphi(\mathbf{x}, t) \in [0,1]$ evolves as
	\begin{equation}
		\frac{\partial \varphi}{\partial t}
		= M\,\nabla^2\!\left[f'(\varphi) - \kappa_{\text{CH}}\nabla^2\varphi\right],
		\label{eq:CH}
	\end{equation}
	where $M$ is the (constant) mobility, $\kappa_{\text{CH}} > 0$ is the
	gradient-energy coefficient that penalises sharp interfaces, and
	$f(\varphi) = \varphi^2(1-\varphi)^2$ is the symmetric double-well free energy.
	Throughout, $\kappa_{\text{CH}}$ denotes the Cahn--Hilliard gradient coefficient
	and is not to be confused with the optical decay constant
	$\kappa = \sqrt{\mu_a/D}$ used from Section~\ref{sec:optics} onwards. The
	spinodal instability condition $f''(\varphi) < 0$ holds for
	$\varphi \in (0.211,\,0.789)$~\cite{Cahn1961}.
	
	Linear stability analysis around a uniform composition $\varphi_0$ gives the
	growth rate of a Fourier mode of wavenumber $k$,
	\begin{equation}
		\sigma(k) = -Mk^2\!\left[f''(\varphi_0) + 2\kappa_{\text{CH}} k^2\right].
		\label{eq:sigma_k}
	\end{equation}
	The fastest-growing mode corresponds to $k^* = k_c/\sqrt{2}$, yielding a dominant
	wavelength
	\begin{equation}
		\lambda^* = 2\pi\sqrt{\frac{2\kappa_{\text{CH}}}{|f''(\varphi_0)|}}.
		\label{eq:lambda_star}
	\end{equation}
	The gradient coefficient is calibrated to target a dominant pore diameter
	$d_{\text{target}}$ via
	\begin{equation}
		\kappa_{\text{CH}} = \frac{1}{2}\left(\frac{d_{\text{target}}}{8\pi}\right)^{\!2}.
		\label{eq:kappa_cal}
	\end{equation}
	The simulation is conducted at $\varphi_{\text{sim}} = 0.5$ (maximum instability,
	$|f''(0.5)| = 1$), and the binary mask is obtained at the end of the run by
	rank-based thresholding of the phase field, which imposes the target porosity
	exactly.
	
	Following Zhu et al.~\cite{Zhu1999}, Eq.~\eqref{eq:CH} is discretised by a
	semi-implicit Fourier spectral scheme,
	\begin{equation}
		\hat{\varphi}^{n+1}_{\mathbf{k}}
		= \frac{\hat{\varphi}^{n}_{\mathbf{k}}
			- \Delta t\,k^2\,\widehat{f'(\varphi^n)}_{\mathbf{k}}}
		{1 + \Delta t\,\kappa_{\text{CH}}\, k^4},
		\label{eq:CH_spectral}
	\end{equation}
	which is unconditionally stable with respect to the linear stabilising term
	(denominator $\geq 1$). All FFTs are performed with the \texttt{cupy.fft} GPU
	backend~\cite{Okuta2017}, with fields stored in double precision.
	
	After the initial decomposition, domains coarsen according to the
	Lifshitz--Slyozov law $L(t) \sim t^{1/3}$~\cite{Lifshitz1961,Wagner1961};
	integration stops when the dominant pore size, monitored through the peak of the
	radial power spectral density, enters a window of width $\sigma$ centred on
	$d_{\text{target}}$. The width parameter $\sigma$ directly controls the breadth
	of the final pore-size distribution: a large $\sigma$ arrests the simulation
	early (young morphology, broad distribution), while a small $\sigma$ allows
	prolonged coarsening (narrow, quasi-monodisperse distribution). The complete set
	of numerical parameters---initial condition, random seed, time step, stopping
	criterion and binarisation rule---is given in \ref{app:CH}, and grid
	convergence is documented in Section~S4 of the Supplementary Information.
	
	\paragraph{Simulation domain} The domain is
	$(N_z, N_y, N_x) = (384, 1024, 1024)$ voxels with
	$a_{\text{vox}} = 1.45$~$\mu$m, giving a physical depth $L = 0.557$~mm and a
	lateral extent of 1.485~mm. Two porosities are studied, $\phi = 0.70$ and
	$\phi = 0.90$; the former is used as the benchmark throughout the main text.
	A three-dimensional isosurface rendering of the solid phase is shown in Fig.~S1
	of the Supplementary Information. Geometry and effective-medium optical
	parameters are summarised in Table~\ref{tab:geometry}.
	
	\begin{table}[htbp]
		\centering
		\caption{Spinodal geometry and effective-medium (EMA) optical parameters for the
			benchmark simulation ($384 \times 1024 \times 1024$ domain, $\phi = 0.70$,
			$\lambda = 365$~nm).}
		\label{tab:geometry}
		\begin{tabular}{lll}
			\toprule
			Parameter & Value & Description \\
			\midrule
			$N_z \times N_y \times N_x$ & $384 \times 1024 \times 1024$ & domain size (voxels) \\
			$a_{\text{vox}}$ & 1.45~$\mu$m & voxel edge length \\
			$\phi$ & 0.7000 & achieved porosity \\
			$d_{\text{peak}}$ & 19.3~$\mu$m & dominant pore diameter \\
			$L$ & 0.557~mm & slab thickness \\
			$\mu_{a,\text{eff}} = \mu_a(1-\phi)$ & 1548~m$^{-1}$ & EMA absorption coefficient \\
			$\mu_{s,\text{eff}}' = \mu_s'(1-\phi)$ & 6035~m$^{-1}$ & EMA reduced scattering \\
			$\ell_{\text{eff}}^* = 1/\mu_{s,\text{eff}}'$ & 165.7~$\mu$m & EMA transport MFP \\
			$D_{\text{eff}} = 1/[3(\mu_{a,\text{eff}}+\mu_{s,\text{eff}}')]$
			& $4.40 \times 10^{-5}$~m & diffusion coefficient \\
			$\kappa_{\text{eff}} = \sqrt{\mu_{a,\text{eff}}/D_{\text{eff}}}$
			& 5934~m$^{-1}$ & EMA decay constant \\
			$\kappa_{\text{eff}} L$ & 3.30 & dimensionless optical thickness \\
			$S_{\text{acc}}$ & $3.10 \times 10^{7}$~m$^{-1}$ & catalytic surface density (total volume) \\
			$S_{\text{acc}}^{\text{solid}} = S_{\text{acc}}/(1-\phi)$ & $1.034 \times 10^{8}$~m$^{-1}$ & catalytic surface density (solid volume) \\
			\bottomrule
		\end{tabular}
	\end{table}
	
	\subsection{Optical properties of the SiO$_2$/TiO$_2$ composite shell}
	\label{sec:optics}
	
	The composite shell (silica matrix $n_{\text{SiO}_2} = 1.45$; anatase TiO$_2$ NPs
	of radius $a_{\text{np}} = 7.25$~nm, volume fraction $f_{\text{NP}} = 0.25$) is
	characterised as follows.
	
	\subsubsection{Maxwell--Garnett effective medium and its justification}
	
	The effective permittivity of the composite is~\cite{Garnett1904}
	\begin{equation}
		\varepsilon_{\text{eff}}
		= \varepsilon_{\text{SiO}_2}\,
		\frac{1 + 2f_{\text{NP}}\,\beta}{1 - f_{\text{NP}}\,\beta},
		\quad
		\beta = \frac{\varepsilon_{\text{TiO}_2} - \varepsilon_{\text{SiO}_2}}
		{\varepsilon_{\text{TiO}_2} + 2\,\varepsilon_{\text{SiO}_2}},
		\label{eq:MG}
	\end{equation}
	with $\varepsilon_{\text{TiO}_2}(\lambda)$ taken from tabulated anatase optical
	constants~\cite{AnataseOptical}. The absorption coefficient of the shell is
	\begin{equation}
		\mu_a = \frac{4\pi\,k_{\text{eff}}}{\lambda},
		\qquad k_{\text{eff}} = \text{Im}(\tilde{n}_{\text{eff}}).
		\label{eq:mua_shell}
	\end{equation}
	
	Maxwell--Garnett (MG) treats one phase as isolated inclusions embedded in a
	continuous host. This asymmetry matches the actual microstructure: discrete
	anatase nanoparticles dispersed in a continuous silica matrix, with
	$f_{\text{NP}} = 0.25$ below the continuum percolation threshold of randomly
	placed spheres ($\approx 0.29$), so that the TiO$_2$ phase does not form a
	connected network. The quasistatic dipole limit is comfortably satisfied, since
	the size parameter is $x = 2\pi n_{\text{SiO}_2} a_{\text{np}}/\lambda \approx
	0.18 \ll 1$. The symmetric Bruggeman (BR) mixing rule~\cite{Bruggeman1935}, which
	treats both phases as interpenetrating aggregates, is topologically less
	appropriate here but constitutes the natural alternative bound; a complete
	sensitivity analysis, including a full 3D MC re-run with BR coefficients, is
	reported in Section~S5 of the Supplementary Information and summarised in
	Section~\ref{sec:uncertainty}.
	
	\subsubsection{Rayleigh scattering}
	
	At $\lambda = 365$~nm the Rayleigh regime applies with isotropic scattering
	($g = 0$, hence $\mu_s' = \mu_s$):
	\begin{equation}
		\mu_s = \frac{f_{\text{NP}}}{V_{\text{NP}}}\,
		\frac{k_m^4\,|\alpha_{\text{CM}}|^2\,V_{\text{NP}}^2}{6\pi},
		\label{eq:mus_rayleigh}
	\end{equation}
	where $k_m = 2\pi n_{\text{SiO}_2}/\lambda$ is the wave vector in the matrix,
	$\alpha_{\text{CM}} = 3(\varepsilon_{\text{TiO}_2} - \varepsilon_{\text{SiO}_2})/
	(\varepsilon_{\text{TiO}_2} + 2\varepsilon_{\text{SiO}_2})$ is the
	Clausius--Mossotti polarisability, and
	$V_{\text{NP}} = \frac{4}{3}\pi a_{\text{np}}^3$. This independent-scatterer
	estimate neglects dependent-scattering corrections, which are not negligible at
	$f_{\text{NP}} = 0.25$. Since $\mu_s'$ enters the MC and the DA identically, the
	estimator comparisons that constitute the conclusions of this work are unaffected
	by this approximation; only the absolute fluence level would shift. A
	$\pm 20$\,\% bracket on $\mu_s'$ is reported in Section~S5 of the Supplementary
	Information, propagated into the absolute uncertainty of
	$k_{\text{surf}}\cdot K_{\text{ads}}$ in Section~\ref{sec:uncertainty},
	and discussed for the reflectance comparison in
	Section~\ref{sec:experimental_optics}.
	
	\subsubsection{Effective-medium domain coefficients}
	
	The MC simulation resolves pore and solid voxels explicitly. The DA, by
	construction, requires homogenised coefficients scaled by the solid
	fraction~\cite{Ishimaru1978}:
	\begin{equation}
		\mu_{a,\text{eff}} = \mu_a\,(1-\phi),
		\qquad
		\mu_{s,\text{eff}}' = \mu_s'\,(1-\phi).
		\label{eq:EMA_scaling}
	\end{equation}
	Both coefficients must be scaled consistently; using the shell $\mu_s'$ together
	with the EMA $\mu_a$ overestimates the decay constant by $\approx 69$\%, as
	quantified in Section~\ref{sec:EMA_consistency}. All coefficients at 365~nm are
	given in Table~\ref{tab:optics}.
	
	\begin{table}[htbp]
		\centering
		\caption{Optical parameters at $\lambda = 365$~nm, $f_{\text{NP}} = 0.25$,
			$a_{\text{np}} = 7.25$~nm, $\phi = 0.70$. Shell values enter the MC property
			table; EMA domain values are used for the DA model.}
		\label{tab:optics}
		\begin{tabular}{llll}
			\toprule
			Parameter & Shell (solid) & EMA domain & Unit \\
			\midrule
			$\mu_a$ & 5159 & 1548 & m$^{-1}$ \\
			$\mu_s'$ & 20118 & 6035 & m$^{-1}$ \\
			$g$ & 0 & 0 & -- \\
			$\ell^*$ & 49.7 & 165.7 & $\mu$m \\
			$\mu_a/\mu_s'$ & 0.256 & 0.256 & -- \\
			$D$ & $1.32 \times 10^{-5}$ & $4.40 \times 10^{-5}$ & m \\
			$\kappa$ & 19779 & 5934 & m$^{-1}$ \\
			$z_b = \frac{2}{3}\ell^*$ & 33.1 & 110.5 & $\mu$m \\
			$n_{\text{eff}}$ & 1.746 & -- & -- \\
			$S_0 = I_0/h\nu$ & \multicolumn{2}{l}{$8.27 \times 10^{19}$}
			& photon\,m$^{-2}$\,s$^{-1}$ \\
			\bottomrule
		\end{tabular}
	\end{table}
	
	\subsection{Analytical diffusion-approximation benchmark}
	\label{sec:DA}
	
	For a collimated planar source at $z = 0$, the fluence decomposes into ballistic
	and diffuse terms~\cite{Ishimaru1978}:
	\begin{equation}
		\Phi_{\text{DA}}(z) = S_0\,e^{-\mu_t z} + \Phi_{\text{diff}}(z),
		\label{eq:DA_total}
	\end{equation}
	with $\mu_t = \mu_{a,\text{eff}} + \mu_{s,\text{eff}}'$. The diffuse term
	satisfies
	\begin{equation}
		D_{\text{eff}}\,\frac{\mathrm{d}^2\Phi_{\text{diff}}}{\mathrm{d}z^2}
		- \mu_{a,\text{eff}}\,\Phi_{\text{diff}}
		= -\mu_{s,\text{eff}}'\,S_0\,e^{-\mu_t z},
		\label{eq:DA_diff_eq}
	\end{equation}
	whose particular solution is $A\,e^{-\mu_t z}$ with
	$A = \mu_{s,\text{eff}}'\,S_0 / [D_{\text{eff}}(\kappa_{\text{eff}}^2 -
	\mu_t^2)]$. Applying extrapolated Robin boundary
	conditions~\cite{Haskell1994} $\Phi_{\text{diff}}(-z_b) =
	\Phi_{\text{diff}}(L+z_b) = 0$ gives
	\begin{equation}
		C_+ = -\frac{A\,e^{-\mu_t(L+z_b)}}{\sinh(\kappa_{\text{eff}} L_e)},
		\qquad
		C_- = -\frac{A\,e^{+\mu_t z_b}}{\sinh(\kappa_{\text{eff}} L_e)},
		\label{eq:Cpm}
	\end{equation}
	with $L_e = L + 2z_b$, and the complete diffuse solution
	\begin{equation}
		\Phi_{\text{diff}}(z) = A\,e^{-\mu_t z}
		+ C_+\sinh[\kappa_{\text{eff}}(z + z_b)]
		+ C_-\sinh[\kappa_{\text{eff}}(L + z_b - z)].
		\label{eq:Phi_diff}
	\end{equation}
	This collimated-source model replaces the isotropic-source Green's function used
	in Ref.~\cite{Vallee2026}, which produces a spurious internal peak at
	$z = \ell^*$ inconsistent with planar illumination; the two are compared directly
	in Fig.~S2 of the Supplementary Information. The effective optical length and the
	depth-averaged fluence follow as
	\begin{equation}
		L_{\text{opt}} = \frac{1}{S_0}\int_0^L \Phi_{\text{DA}}(z)\,\mathrm{d}z,
		\qquad
		\langle\Phi\rangle_{\text{DA}} = \frac{S_0\,L_{\text{opt}}}{L}.
		\label{eq:Lopt_DA}
	\end{equation}
	
	\subsection{Three-dimensional Monte Carlo simulation}
	\label{sec:MC}
	
	Photon transport is simulated with the GPU Monte Carlo code
	MCX~\cite{Fang2009,Yu2018} through its Python interface. Pore voxels carry
	$\mu_a = \mu_s = 0$; solid voxels carry the shell coefficients of
	Table~\ref{tab:optics}. The complete solver configuration is listed in
	\ref{app:MC}.
	
	\subsubsection{Source and boundary conditions}
	
	A planar collimated source illuminates the full $z = 0$ face. The four lateral
	faces carry specular mirrors and the two axial faces are open, so that photons
	crossing $z < 0$ or $z > L$ leave the domain and are recorded
	(\texttt{bc = "mmamma"}). Mirror lateral boundaries emulate an infinite slab
	illuminated uniformly, which is the physically intended configuration, and avoid
	both the non-physical lateral flux sink of absorbing boundaries and the
	artificial upturn of $\Phi(z)$ at depth produced by cyclic boundaries when
	$L_{\text{lat}}/\ell_{\text{eff}}^* \lesssim 10$ (here 8.9). A one-voxel air pad
	is added on both axial faces so that the air/solid interface lies inside the
	computational volume; the entry Fresnel reflection is then captured natively by
	the internal Fresnel treatment (\texttt{isreflect = 1}) rather than through a
	separate specular term.
	
	\subsubsection{Photon budget, time gate and statistics}
	
	The number of launched packets is set by
	$N = \max\!\left[10\,N_xN_y\,e^{\mu_{a,\text{shell}}L},\,10^6\right]
	= 1.85\times10^{8}$, and the random seed is fixed and reported
	(\ref{app:MC}). The solver uses the standard weighted-photon
	(implicit-capture) scheme with Russian-roulette termination, which is its
	variance-reduction mechanism. The simulation time gate is diffusion-aware,
	\begin{equation}
		t_{\text{end}} = \max\!\left[\frac{5\,n_{\text{eff}}}{\mu_{a,\text{shell}}\,c},\;
		\frac{6L^2}{D_{\text{phys}}}\right],
		\qquad D_{\text{phys}} = \frac{c/n_{\text{eff}}}{3(\mu_{a,\text{eff}}+\mu_{s,\text{eff}}')},
		\label{eq:tend}
	\end{equation}
	giving $2.46 \times 10^{-10}$~s at $\phi = 0.70$ and $8.21 \times 10^{-11}$~s at
	$\phi = 0.90$. The diffusive floor guarantees complete collection of the
	late-exiting diffuse reflectance, which a purely absorption-based gate would
	truncate. Statistical convergence, morphological seed-to-seed variability and
	grid convergence are documented in Section~S4 of the Supplementary Information.
	
	\subsubsection{Photon budget closure}
	
	Absorptance is obtained from the absorbed-energy counter and reflectance and
	transmittance from the boundary escape map, independently of one another:
	\begin{equation}
		A = \frac{E_{\text{abs}}}{E_{\text{tot}}},
		\qquad
		R = \frac{a_{\text{vox}}^2\,t_{\text{end}}}{1}\sum_{x,y} \Xi(x,y,0),
		\qquad
		T = \frac{a_{\text{vox}}^2\,t_{\text{end}}}{1}\sum_{x,y} \Xi(x,y,N_z+1),
		\label{eq:RTA}
	\end{equation}
	where $\Xi$ is the escape-weight map recorded on the two air-pad planes and
	$E_{\text{tot}}$ the total launched weight. Because none of the three channels is
	obtained by difference, the closure
	\begin{equation}
		R + T + A = 1
		\label{eq:closure}
	\end{equation}
	is a \emph{measurement} and therefore a genuine per-run validation. All nine
	production and auxiliary simulations reported in this work satisfy
	Eq.~\eqref{eq:closure} between $0.99999994$ and $1.00000006$.
	
	\subsubsection{Fluence normalisation}
	
	The raw solver output is converted to absolute physical fluence via
	\begin{equation}
		\Phi(x,y,z) = \Phi_{\text{MC}}(x,y,z) \times S_0 \times A_{\text{source}}
		\times t_{\text{end}},
		\label{eq:Phi_norm}
	\end{equation}
	with $A_{\text{source}} = N_xN_y a_{\text{vox}}^2$ the illuminated area.
	
	\subsection{Fluence estimators and the phase-contrast identity}
	\label{sec:estimators}
	
	Because photocatalysis occurs exclusively in the solid phase, the physically
	relevant fluence average weights by the solid indicator
	$s(\mathbf{r}) = 1 - m(\mathbf{r})$, where $m = 1$ in pores and $0$ in solid:
	\begin{equation}
		\langle\Phi\rangle_{\text{MC,3D}}
		= \frac{\sum_{\mathbf{r}} \Phi(\mathbf{r})\,s(\mathbf{r})}
		{\sum_{\mathbf{r}} s(\mathbf{r})}.
		\label{eq:Phi_MC3D}
	\end{equation}
	This is contrasted with the conventional volume average
	\begin{equation}
		\langle\Phi\rangle_{\text{MC,vol}}
		= \frac{1}{N_x N_y L}\int_0^L \sum_{x,y} \Phi(x,y,z)\,\mathrm{d}z,
		\label{eq:Phi_MCvol}
	\end{equation}
	obtained by spatial reduction of the same three-dimensional Monte Carlo
	fluence field over all voxels (solid and pore). It is not the output of a
	separate one-dimensional Monte Carlo simulation; the companion paper's 1D MC
	benchmark~\cite{Vallee2026} is a different object, computed on a homogeneous
	slab. The volume average mixes the solid-phase signal with the pore voxels. Denoting by
	$\langle\Phi\rangle_{\text{pore}}$ the corresponding pore-weighted average, the
	volume average decomposes exactly as
	$\langle\Phi\rangle_{\text{MC,vol}} = \phi\,\langle\Phi\rangle_{\text{pore}} +
	(1-\phi)\,\langle\Phi\rangle_{\text{MC,3D}}$, so that, introducing the
	\emph{phase-contrast ratio}
	\begin{equation}
		r \equiv \frac{\langle\Phi\rangle_{\text{pore}}}{\langle\Phi\rangle_{\text{MC,3D}}},
		\label{eq:r_def}
	\end{equation}
	the concentration ratio between the two MC estimators obeys the identity
	\begin{equation}
		\boxed{
			\frac{\langle\Phi\rangle_{\text{MC,3D}}}{\langle\Phi\rangle_{\text{MC,vol}}}
			= \frac{1}{\phi\,r + (1-\phi)} .
		}
		\label{eq:dilution}
	\end{equation}
	Equation~\eqref{eq:dilution} is exact and contains no adjustable parameter. It
	separates cleanly what is optical ($r$) from what is compositional ($\phi$), and
	it is the quantitative backbone of the mechanism established in
	Section~\ref{sec:mechanism}. Since
	$k_{\text{surf}} \cdot K_{\text{ads}} \propto 1/\langle\Phi\rangle$
	(Eq.~\eqref{eq:ksKads}), the choice of estimator directly controls the extracted
	kinetic descriptor.
	
	\subsection{Discrepancy metric}
	\label{sec:metric}
	
	Comparisons between MC and DA depth profiles are quantified by a
	fluence-weighted relative root-mean-square deviation,
	\begin{equation}
		\mathrm{RMS}_w = \left[\frac{\sum_z \Phi_{\text{DA}}(z)\,
			\bigl(\Phi_{\text{MC}}(z)-\Phi_{\text{DA}}(z)\bigr)^2/\Phi_{\text{DA}}^2(z)}
		{\sum_z \Phi_{\text{DA}}(z)}\right]^{1/2}.
		\label{eq:RMSw}
	\end{equation}
	The unweighted relative RMS, obtained by setting the weights to unity, is
	dominated by the deep tail where both profiles approach zero and relative errors
	diverge; it is therefore a poor measure of how well the DA reproduces the photon
	budget, and for strongly attenuating slabs it can even reverse the ordering of
	configurations. We use Eq.~\eqref{eq:RMSw} as the primary metric throughout and
	report the unweighted and log-ratio variants alongside it in Table~S1 of the
	Supplementary Information. All RMS values quoted in the text, in the figure
	panels and in the captions are the fluence-weighted $\mathrm{RMS}_w$ of
	Eq.~\eqref{eq:RMSw} unless explicitly labelled otherwise; for reference, the
	unweighted variant of the $\phi = 0.70$ spinodal benchmark is 70.6\%, which
	illustrates the tail dominance discussed above.
	
	\subsection{Kinetic extraction}
	\label{sec:kinetic_extraction}
	
	The catalytic surface density is given by Eq.~\eqref{eq:Sacc} and the intrinsic
	volumetric rate constant by Eq.~\eqref{eq:kV}. From the measured
	$k_{\text{app,exp}} = 0.0052$~min$^{-1} = 8.67\times10^{-5}$~s$^{-1}$,
	\begin{equation}
		k_{\text{surf}} \cdot K_{\text{ads}}
		= \frac{k_{\text{app,exp}}}
		{S_{\text{acc}}^{\text{solid}}\,\phi_{\text{int}}\,\mu_{a,\text{eff}}\,
			\langle\Phi\rangle},
		\label{eq:ksKads_extract}
	\end{equation}
	performed for each of the three estimators
	$\langle\Phi\rangle \in \{\langle\Phi\rangle_{\text{MC,3D}},\,
	\langle\Phi\rangle_{\text{MC,vol}},\,\langle\Phi\rangle_{\text{DA}}\}$.
	
	The pairing of $S_{\text{acc}}^{\text{solid}}$ with $\mu_{a,\text{eff}}$ in
	Eq.~\eqref{eq:ksKads_extract} follows directly from the domain-averaging
	identity established in Section~\ref{sec:kinetic_chain}
	(Eq.~\eqref{eq:rVmean_norm}), which shows $S_{\text{acc}}^{\text{solid}}\,
	\mu_{a,\text{eff}} \equiv S_{\text{acc}}\,\mu_a$; pairing $S_{\text{acc}}$ with
	$\mu_{a,\text{eff}}$ instead would apply the solid-fraction factor $(1-\phi)$
	twice. Equation~\eqref{eq:ksKads_extract} applies the same normalisation to
	all three estimators, so the inter-estimator ratios reported below are
	independent of this choice.
	
	Under the present assumptions, at the upper bound
	$k_{\text{surf}} = a_{\text{np}}$, Eq.~\eqref{eq:rVmean_norm} gives
	$r_{V,\text{mean}}/G_{\text{mean}} = 3 f_{\text{NP}} = 0.75$ at both porosities,
	with $G_{\text{mean}} = \phi_{\text{int}}\,\mu_{a,\text{eff}}\,
	\langle\Phi\rangle_{\text{MC,3D}}$ the mean carrier generation rate of
	Eq.~\eqref{eq:G}: an interpretable capture efficiency, below unity and
	independent of porosity under the present assumptions. This ratio follows from
	the assumed equations and is not an independent control.
	
	\section{Results}
	\label{sec:results}
	
	\subsection{Spinodal morphology and pore-size distribution}
	\label{sec:morphology}
	
\begin{figure}[htbp]
		\centering
		\includegraphics[width=\textwidth]{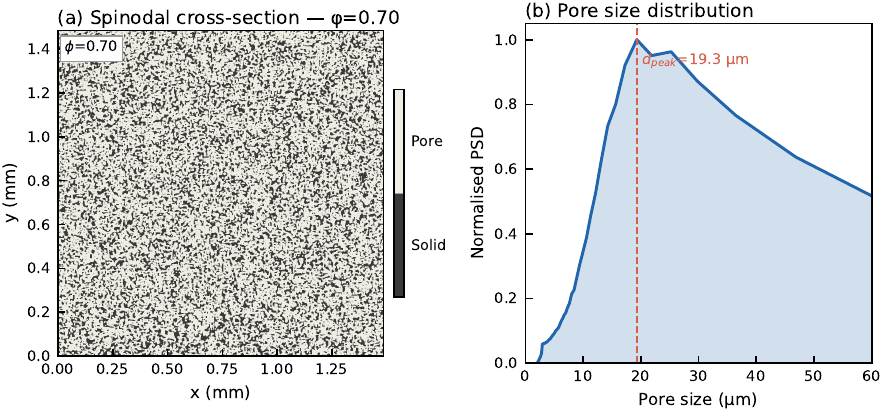}
		\caption{Spinodal morphology of the $384 \times 1024 \times 1024$ aerogel domain
			($\phi = 0.70$, $a_{\text{vox}} = 1.45$~$\mu$m). (a)~Mid-plane ($z = N_z/2$)
			cross-section of the binary mask; light grey: pore, dark: solid. The bicontinuous
			interpenetration of both phases is characteristic of spinodal decomposition and
			consistent with the Si(HIPE) aerogel microstructure of Ref.~\cite{Vardon2025}.
			(b)~Normalised radial power spectral density. The dominant peak at
			$d_{\text{peak}} = 19.3$~$\mu$m (red dashed line) agrees with the target
			17.4~$\mu$m to within 11\%. The secondary shoulder near $\approx 37$~$\mu$m
			arises from partially coarsened pores.}
		\label{fig:morphology}
	\end{figure}

	Figure~\ref{fig:morphology} shows the $384 \times 1024 \times 1024$ spinodal
	geometry. Panel~(a) displays the mid-plane cross-section at porosity
	$\phi = 0.70$, with the characteristic bicontinuous solid/pore architecture of
	spinodal decomposition. Panel~(b) shows the radial power spectral density (PSD),
	with a dominant peak at $d_{\text{peak}} = 19.3$~$\mu$m, close to the target
	17.4~$\mu$m, and a secondary shoulder at $\approx 37$~$\mu$m---twice the dominant
	scale---reflecting partial Lifshitz--Slyozov coarsening halted by the stopping
	criterion~\cite{Lifshitz1961}.
	
	These targets are directly traceable to the experimental characterisation of the
	parent Si(HIPE) system. The simulated dominant cavity diameter agrees to within
	7\% with the SEM macropore-size histogram of Si(HIPE)1000, log-normal with peak
	$D^{*}_{\text{macro}} = 18.0$~$\mu$m and width parameter
	$\sigma = 0.34$~\cite{Vardon2025}, and both distributions share a log-normal-like
	shape with a coarse tail. The interconnection scales are likewise consistent:
	mercury intrusion on the TiO$_2$-loaded monoliths gives junction-window apertures
	of 1--10~$\mu$m~\cite{Layan2025} and image analysis a throat surface fraction of
	55\% for Si(HIPE)1000~\cite{Vardon2025}, in line with a bicontinuous architecture
	whose cavity diameters exceed its windows. The chord-length analysis of
	Section~\ref{sec:mechanism} confirms this asymmetry quantitatively for the
	simulated mask, with a mean solid-ligament chord of 8.1~$\mu$m against a mean
	pore chord of 18.6~$\mu$m.
	
	\subsection{Photon energy balance}
	\label{sec:energy}
	
	Table~\ref{tab:energy} summarises the photon energy balance for both porosities.
	At $\phi = 0.70$ the absorptance $A = 0.643$ confirms that 64\% of the incident
	UV photons are absorbed in the TiO$_2$-bearing solid and are available for
	photocatalysis, while $R = 0.354$ is diffuse reflectance returned through the
	illuminated face and $T = 0.003$ is genuine forward transmission. The front/back
	fluence ratio of 318 shows that the domain is effectively opaque at this
	porosity. The budget closes to $R+T+A = 1.0000$, which, since the three channels
	are measured independently (Section~\ref{sec:MC}), validates the tally.
	
	Raising the porosity to $\phi = 0.90$ scales both effective coefficients by
	$(1-\phi)$ and reduces the optical thickness to $\kappa_{\text{eff}} L = 1.10$.
	Absorptance then falls to 0.430 while reflectance rises to 0.518 and
	transmittance to 0.052; the front/back ratio drops to 25. A larger fraction of the incident photons leaves the monolith unabsorbed, so the
	absorptance falls from 0.643 to 0.430 and the irradiated depth increases
	(front/back ratio 318 to 25). Per unit of catalytic solid, however, the
	absorbed-photon rate increases, because $A/(1-\phi)$ roughly doubles, in line
	with the higher solid-phase fluence. Both reflectance
	values are confronted with the UV--vis diffuse-reflectance measurements on the
	companion monolith series in Section~\ref{sec:experimental_optics}.
	
	\begin{table}[htbp]
		\centering
		\caption{Photon energy balance from the 3D MC simulations
			($384 \times 1024 \times 1024$, $N = 1.85 \times 10^8$ packets,
			\texttt{bc = "mmamma"}, $\lambda = 365$~nm). $R$, $T$ and $A$ are measured
			independently, so that $R+T+A$ is a validation rather than a normalisation
			(Eq.~\eqref{eq:closure}).}
		\label{tab:energy}
		\begin{tabular}{lccccc}
			\toprule
			Porosity & $\kappa_{\text{eff}}L$ & $R$ & $T$ & $A$ & $R+T+A$ \\
			\midrule
			$\phi = 0.70$ & 3.30 & 0.354 & 0.0031 & 0.643 & 1.0000 \\
			$\phi = 0.90$ & 1.10 & 0.518 & 0.0516 & 0.430 & 1.0000 \\
			\bottomrule
		\end{tabular}
	\end{table}
	
	\subsection{Fluence depth profile: MC versus diffusion approximation}
	\label{sec:profiles}
	
\begin{figure}[htbp]
		\centering
		\includegraphics[width=\textwidth]{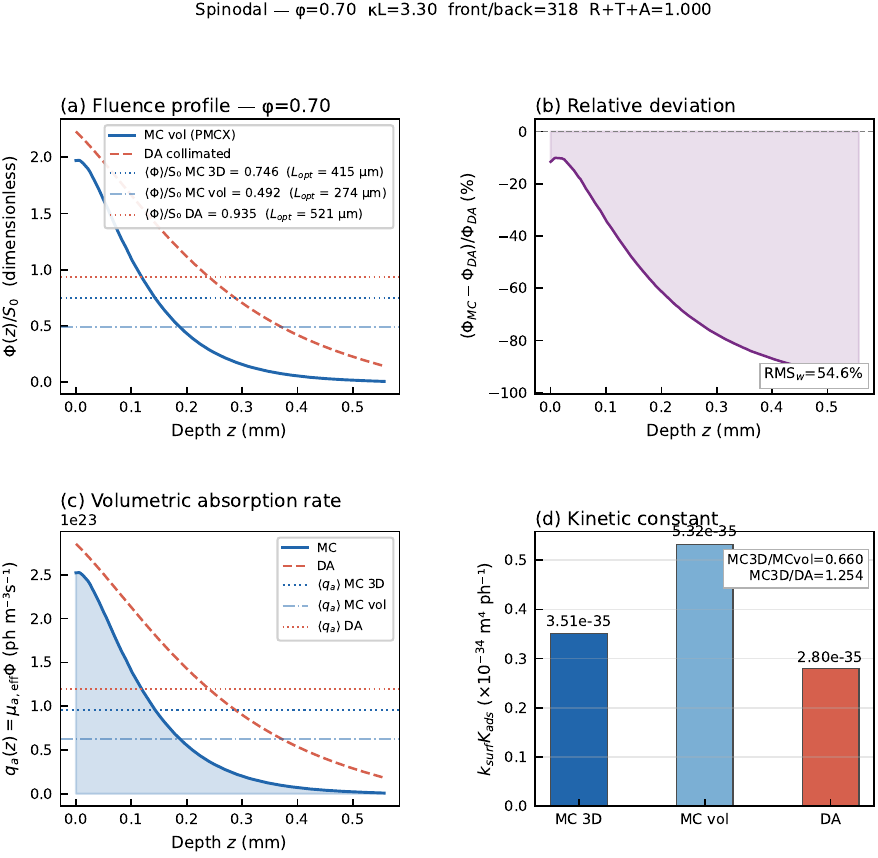}
		\caption{Photon fluence in the $384 \times 1024 \times 1024$ spinodal domain
			($\lambda = 365$~nm, $\phi = 0.70$, $\kappa_{\text{eff}} L = 3.30$, front/back
			$= 318$). (a)~Dimensionless profile $\Phi(z)/S_0$ versus depth: MC lateral
			average (solid blue) and DA
			collimated EMA solution (red dashed, Eqs.~\eqref{eq:DA_total}--\eqref{eq:Phi_diff}).
			Horizontal lines mark the three mean-fluence estimators:
			$\langle\Phi\rangle_{\text{MC,3D}}$ (solid-phase weighted, dark blue dotted),
			$\langle\Phi\rangle_{\text{MC,vol}}$ (volume average, blue dash-dot) and
			$\langle\Phi\rangle_{\text{DA}}$ (red dotted), each labelled by the
			corresponding effective optical length
			$L_{\text{opt}} = \langle\Phi\rangle L/S_0$. (b)~Relative deviation
			$(\Phi_{\text{MC}} - \Phi_{\text{DA}})/\Phi_{\text{DA}}$;
			$\mathrm{RMS}_w = 54.6$\%. (c)~Local absorption rate
			$q_a(z) = \mu_{a,\text{eff}}\,\Phi(z)$ for MC and DA, with mean estimators as
			horizontal lines. (d)~$k_{\text{surf}} \cdot K_{\text{ads}}$ extracted via
			Eq.~\eqref{eq:ksKads_extract} for the three estimators; ratios annotated.}
		\label{fig:fluence_profile}
	\end{figure}

	Figure~\ref{fig:fluence_profile} compares the MC and DA fluence profiles at
	$\phi = 0.70$.
	
	\paragraph{Panel (a) --- absolute profiles} The DA collimated solution shows the
	ballistic entry followed by a smooth diffuse decay at rate $\kappa_{\text{eff}}$.
	The MC profile displays a sharper initial peak at $z \approx 10$~$\mu$m, narrower
	than $\ell_{\text{eff}}^*$ because first-scatter events are localised in solid
	voxels much smaller than the EMA scale. Beyond $z \approx 0.05$~mm the MC falls
	below the DA. The three mean-fluence estimators span a factor of $\approx 1.9$:
	$\langle\Phi\rangle_{\text{MC,3D}} = 6.17 \times 10^{19}$,
	$\langle\Phi\rangle_{\text{MC,vol}} = 4.07 \times 10^{19}$ and
	$\langle\Phi\rangle_{\text{DA}} = 7.73 \times
	10^{19}$~photon\,m$^{-2}$\,s$^{-1}$.
	
	\paragraph{Panel (b) --- relative deviation} The deviation is negative over most
	of the domain. The fluence-weighted RMS is 54.6\%
	(Eq.~\eqref{eq:RMSw}); near the entry face ($z < z_b = 110$~$\mu$m) it falls to
	$\approx -20$\%, confirming reasonable near-field accuracy of the
	collimated-source DA. The origin of the 54.6\% is decomposed in
	Section~\ref{sec:homogeneous}.
	
	\paragraph{Panels (c) and (d)} The local absorption rate
	$q_a(z) = \mu_{a,\text{eff}}\Phi(z)$ mirrors the fluence profile. Panel~(d) shows
	that the three estimators yield $k_{\text{surf}} \cdot K_{\text{ads}}$ values
	differing by up to a factor of 1.9, as analysed in
	Section~\ref{sec:rate_constants}.
	
	\subsection{Two-dimensional fluence map}
	\label{sec:map}
	
\begin{figure}[htbp]
		\centering
		\includegraphics[width=0.55\textwidth]{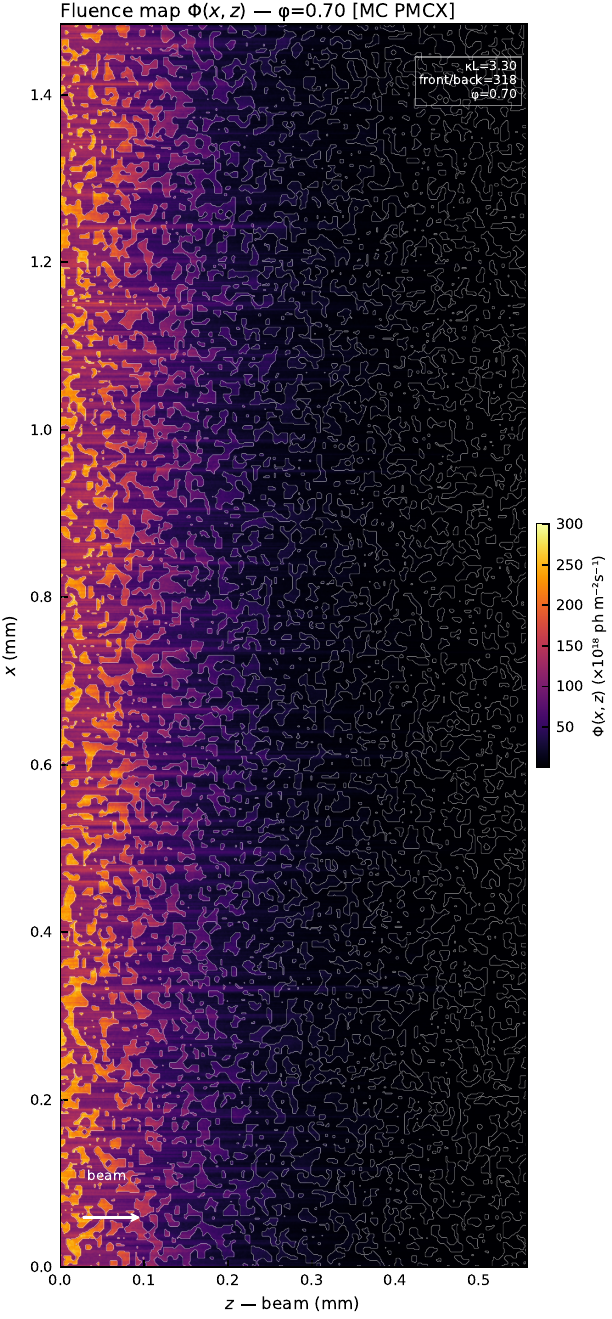}
		\caption{Two-dimensional fluence map $\Phi(x,z)$ at the mid-plane cross-section
			($y = N_y/2$) of the $384 \times 1024 \times 1024$ domain ($\phi = 0.70$,
			$\kappa_{\text{eff}} L = 3.30$, front/back $= 318$). Colormap: inferno
			($\times 10^{18}$~photon\,m$^{-2}$\,s$^{-1}$). White contours: solid/pore
			interface from the binary mask. The beam propagates in the $+z$ direction (arrow,
			lower left). Superimposed on the rapid decay from the illuminated face (left),
			the fluence field is strongly textured: the bright filaments track the solid
			ligaments, in which scattering sustains a high local path density, while the
			non-scattering void channels remain comparatively dark. Depth-normalised
			statistics (Section~\ref{sec:mechanism}) show that 88.5\% of the voxels in the
			brightest fluence decile belong to the solid phase, which occupies 29.9\% of the
			volume.}
		\label{fig:fluence_map}
	\end{figure}

	Figure~\ref{fig:fluence_map} shows the fluence map $\Phi(x,z)$ at the mid-plane
	cross-section. The dominant feature is the rapid decay from the illuminated face:
	beyond $z \approx 0.1$~mm most of the domain is dark, consistent with the
	front/back ratio of 318. Superimposed on that decay is a pronounced lateral
	texture that follows the solid/pore interface contours. The bright filaments
	coincide with the solid ligaments and the dark regions with the void channels;
	the quantitative statistics establishing this correspondence are given in
	Section~\ref{sec:mechanism}. The map thus displays, in real space, the phase
	contrast that Eq.~\eqref{eq:dilution} converts into the concentration ratio.
	
	\subsection{Quantitative signatures of the transport mechanism}
	\label{sec:mechanism}
	
\begin{figure}[htbp]
		\centering
		\includegraphics[width=\textwidth]{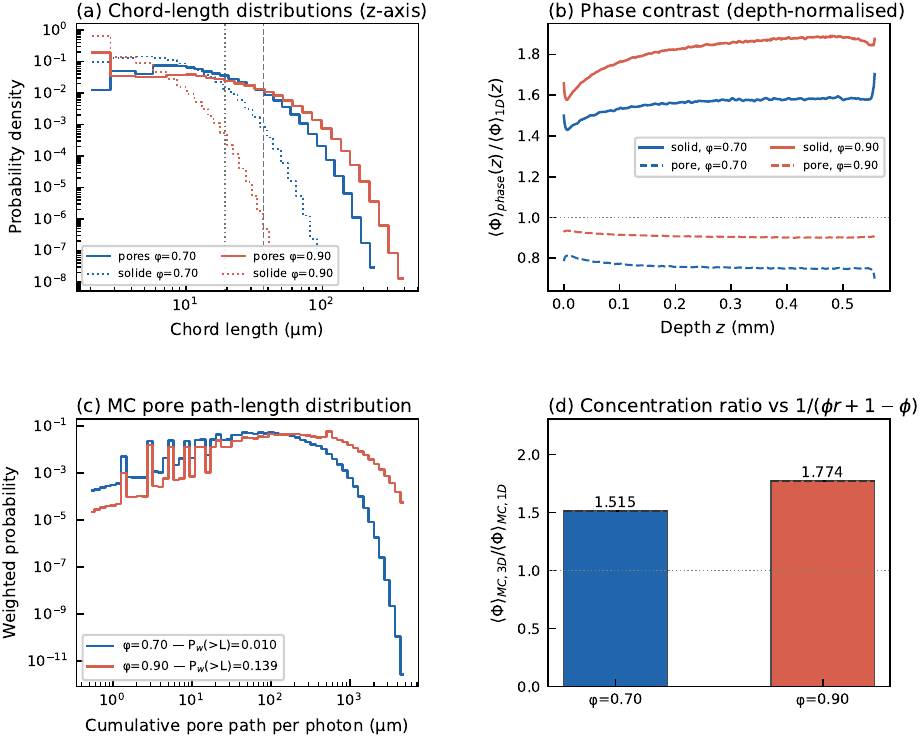}
		\caption{Quantitative signatures of the transport mechanism at $\phi = 0.70$
			(blue) and $\phi = 0.90$ (red). (a)~Chord-length distributions along the beam
			axis for the pore (solid lines) and solid (dotted) phases; vertical grey lines
			mark the primary (19.3~$\mu$m) and secondary (37~$\mu$m) structural scales of the
			PSD (Fig.~\ref{fig:morphology}b). (b)~Phase contrast: depth-resolved fluence of
			each phase normalised by the lateral mean at the same depth,
			$\langle\Phi\rangle_{\text{phase}}(z)/\langle\Phi\rangle_{\text{MC,vol}}(z)$. The
			solid phase lies systematically above unity and the pore phase below, with both
			ratios nearly constant across the slab. (c)~Weighted distribution of the
			cumulative pore path length per escaping photon, from a dedicated detector run;
			$P_w(>L)$ is the weight fraction whose in-pore path exceeds the slab thickness.
			(d)~Morphological concentration ratio
			$\langle\Phi\rangle_{\text{MC,3D}}/\langle\Phi\rangle_{\text{MC,vol}}$ (bars,
			with the seed-to-seed standard deviation of Table~S3 as error bar) compared with
			the dilution identity of Eq.~\eqref{eq:dilution} evaluated with the measured
			phase contrast $r$ (dashed segments).}
		\label{fig:channelling}
	\end{figure}

	Figure~\ref{fig:channelling} gathers four independent quantitative
	characterisations of the transport mechanism, for both porosities.
	
	\paragraph{(a) Chord-length distributions} Computed from the binary masks along
	the beam axis, the pore chord distribution is broad and heavy-tailed. The mean
	pore chord is 18.6~$\mu$m at $\phi = 0.70$ and 22.8~$\mu$m at $\phi = 0.90$,
	against mean solid chords of 8.1 and 2.6~$\mu$m respectively. The length-weighted
	fraction of pore chords exceeding the secondary PSD scale is
	$P_{\text{len}}(\ell > 37~\mu\text{m}) = 0.314$ and $0.557$, and
	$P_{\text{len}}(\ell > 74~\mu\text{m}) = 0.046$ and $0.195$. Long unobstructed
	void segments are therefore geometrically available, increasingly so at high
	porosity.
	
	\paragraph{(b) Phase-resolved fluence} Normalising the fluence at each depth by
	its lateral mean removes the two-decade depth trend and isolates the phase
	contrast. The solid phase carries $1.575 \pm 0.007$ times the local mean and the
	pore phase $0.755 \pm 0.008$ at $\phi = 0.70$ ($1.851 \pm 0.009$ and
	$0.906 \pm 0.003$ at $\phi = 0.90$), and both are essentially depth-independent
	across the slab.
	
	These depth-normalised contrasts weight every depth equally. The phase-contrast
	ratio $r$ of Eq.~\eqref{eq:r_def} is a different quantity: it is the ratio of the
	two \emph{global} fluence averages, and is therefore dominated by the bright
	near-surface region, where the contrast is weaker than in the depth-normalised
	average (Fig.~\ref{fig:channelling}b). The two should not be conflated, and $r$ is
	not obtained by dividing the two numbers just quoted. It follows instead from the
	exact decomposition $\langle\Phi\rangle_{\text{MC,vol}} =
	\phi\,\langle\Phi\rangle_{\text{pore}} + (1-\phi)\,\langle\Phi\rangle_{\text{MC,3D}}$,
	which gives $\langle\Phi\rangle_{\text{pore}} = 3.170 \times 10^{19}$ at
	$\phi = 0.70$ and $6.381 \times 10^{19}$~photon\,m$^{-2}$\,s$^{-1}$ at
	$\phi = 0.90$, hence $r = 0.5142$ and $r = 0.5152$: identical to within 0.2\%
	over the two simulated porosities.
	Substituting into Eq.~\eqref{eq:dilution} returns 1.5153 and 1.7739, recovering
	the simulated concentration ratios by algebraic consistency rather than as an
	independent prediction.
	
	\paragraph{(c) Per-photon path-length statistics} A dedicated MC run with
	detectors on both axial faces records, for every escaping photon, the partial
	path length accumulated in each medium ($10^7$ launched packets). The weighted
	distribution of the cumulative pore path per photon is heavy-tailed: its mean is
	93.8~$\mu$m at $\phi = 0.70$ and 276.5~$\mu$m at $\phi = 0.90$, with weight
	fractions of 0.617 and 0.827 exceeding the secondary pore scale (37~$\mu$m) and
	0.0096 and 0.139 exceeding the full slab thickness $L = 557$~$\mu$m. At the
	higher porosity, one escaping photon in seven has therefore travelled further
	inside the void phase than the sample is thick, which is possible only through
	multiply-connected void channels. This ensemble is exit-weighted and comprises
	the escaping fraction of the budget (38\% and 57\% of the launched weight
	respectively, of which 99\% and 91\% leaves through the front face); it
	characterises path topology and not the fluence budget itself.
	
	\paragraph{(d) Concentration ratio versus the dilution prediction} The measured
	concentration ratios, $1.515$ and $1.774$, are compared with
	Eq.~\eqref{eq:dilution} evaluated with the measured $r$. The agreement is exact by construction of $r$, which is the point: reproducing
	the concentration ratio from the measured $r$ is algebraic consistency, not an
	independent prediction. A single measured optical contrast, common to both
	porosities, together with the composition, accounts for the entire morphological
	concentration effect in these two geometries.
	
	\paragraph{Co-localisation statistics} On deep cross-sections
	($z \ge 0.1$~mm), and after removal of the depth trend, $88.5 \pm 3.2$\% of the
	voxels in the brightest fluence decile lie in the solid phase, against a solid
	volume fraction of 29.9\% ($80.2 \pm 1.8$\% versus 10.0\% at $\phi = 0.90$).
	Within the pore phase, the Spearman correlation between local fluence and local
	pore diameter, the latter measured by a Euclidean distance transform, is
	$\rho = +0.034 \pm 0.008$ ($+0.031 \pm 0.005$ at $\phi = 0.90$): resolvable given
	the sample size, but negligible in magnitude. Local fluence is therefore
	controlled by which phase a voxel belongs to, and not by the size of the void it
	sits in.
	
	\subsection{Photocatalytic rate constants}
	\label{sec:rate_constants}
	
	Table~\ref{tab:kinetics} compares the kinetic quantities obtained from the three
	estimators at both porosities.
	
	\begin{table}[htbp]
		\centering
		\caption{Photocatalytic quantities from the three fluence estimators at
			$\lambda = 365$~nm, $\phi_{\text{int}} = 0.25$,
			$k_{\text{surf}} = 7.25 \times 10^{-9}$~m,
			$k_{\text{app,exp}} = 0.0052$~min$^{-1}$; domain $384 \times 1024 \times 1024$,
			\texttt{bc = "mmamma"}. The MC~3D solid-weighted estimator
			(Eq.~\eqref{eq:Phi_MC3D}) is the highest-fidelity reference considered in this
			study, within the present model assumptions.
			$L_{\text{opt}} = \langle\Phi\rangle L / S_0$. The descriptor
			$k_{\text{surf}} \cdot K_{\text{ads}}$ is extracted with
			Eq.~\eqref{eq:ksKads_extract}, i.e.\ with the solid-referenced surface density
			$S_{\text{acc}}^{\text{solid}}$ paired with $\mu_{a,\text{eff}}$
			(Eq.~\eqref{eq:rVmean_norm}).}
		\label{tab:kinetics}
		\begin{tabular}{lccc}
			\toprule
			Quantity & MC~3D (solid) & MC~vol (volume) & DA (collimated) \\
			\midrule
			\multicolumn{4}{l}{\textbf{$\phi = 0.70$} ($\kappa_{\text{eff}}L = 3.30$, $S_{\text{acc}} = 3.10\times10^{7}$~m$^{-1}$, $S_{\text{acc}}^{\text{solid}} = 1.034\times10^{8}$~m$^{-1}$)} \\
			$\langle\Phi\rangle$ (photon\,m$^{-2}$\,s$^{-1}$)
			& $6.17 \times 10^{19}$ & $4.07 \times 10^{19}$ & $7.73 \times 10^{19}$ \\
			$L_{\text{opt}}$ ($\mu$m) & 415 & 274 & 521 \\
			$r_{V,\text{mean}}$ (photon\,m$^{-3}$\,s$^{-1}$)
			& $1.789 \times 10^{22}$ & $1.181 \times 10^{22}$ & $2.244 \times 10^{22}$ \\
			$k_{\text{surf}} \cdot K_{\text{ads}}$ (m$^4$\,photon$^{-1}$)
			& $3.512 \times 10^{-35}$ & $5.321 \times 10^{-35}$ & $2.800 \times 10^{-35}$ \\
			Ratio (rel.\ to MC~3D) & 1.000 & 1.515 & 0.797 \\
			\midrule
			\multicolumn{4}{l}{\textbf{$\phi = 0.90$} ($\kappa_{\text{eff}}L = 1.10$, $S_{\text{acc}} = 1.03\times10^{7}$~m$^{-1}$, $S_{\text{acc}}^{\text{solid}} = 1.034\times10^{8}$~m$^{-1}$)} \\
			$\langle\Phi\rangle$ (photon\,m$^{-2}$\,s$^{-1}$)
			& $1.238 \times 10^{20}$ & $6.98 \times 10^{19}$ & $1.266 \times 10^{20}$ \\
			$L_{\text{opt}}$ ($\mu$m) & 834 & 470 & 852 \\
			$r_{V,\text{mean}}$ (photon\,m$^{-3}$\,s$^{-1}$)
			& $1.198 \times 10^{22}$ & $6.753 \times 10^{21}$ & $1.225 \times 10^{22}$ \\
			$k_{\text{surf}} \cdot K_{\text{ads}}$ (m$^4$\,photon$^{-1}$)
			& $5.245 \times 10^{-35}$ & $9.304 \times 10^{-35}$ & $5.131 \times 10^{-35}$ \\
			Ratio (rel.\ to MC~3D) & 1.000 & 1.774 & 0.978 \\
			\bottomrule
		\end{tabular}
	\end{table}
	
	\paragraph{Morphological concentration effect} The ratio
	$\langle\Phi\rangle_{\text{MC,3D}}/\langle\Phi\rangle_{\text{MC,vol}} = 1.515$ at
	$\phi = 0.70$ means that the catalytic solid phase experiences 51.5\% more
	photons than the volume average suggests; at $\phi = 0.90$ the excess reaches
	77.4\%. Section~\ref{sec:mechanism} shows that both values follow from
	Eq.~\eqref{eq:dilution} with the same optical contrast.
	
	\paragraph{Discrepancy between estimators} At $\phi = 0.70$ the three estimators
	give $k_{\text{surf}} \cdot K_{\text{ads}}$ values spanning a factor of 1.9. The
	volume average overestimates the descriptor by 51.5\%: diluting the solid-phase
	signal with dark pore voxels underestimates $\langle\Phi\rangle$, and a larger
	intrinsic reactivity is then required to reproduce $k_{\text{app,exp}}$. The DA
	underestimates it by 20\%, because it overestimates $\langle\Phi\rangle$: it
	cannot represent the discrete two-phase structure, and it is applied at the edge
	of its validity domain (Section~\ref{sec:DA_validity}). The MC~3D solid-weighted
	value, $3.512 \times 10^{-35}$~m$^4$\,photon$^{-1}$, is the spatially resolved
	reference within the present model assumptions. The two error sources are
	independent: the ratio MC~3D/DA $= 1.254$
	measures the failure of the effective-medium DA, whereas MC~3D/MC~vol $= 0.660$
	measures the averaging artefact of the volume estimator. Their partial
	cancellation in the DA/MC~vol ratio ($0.526$) is accidental and should not be
	relied upon.
	
	\paragraph{Sensitivity to $\phi_{\text{int}}$.} Since
	$k_{\text{surf}} \cdot K_{\text{ads}} \propto 1/\phi_{\text{int}}$, the three
	estimator values scale identically with $\phi_{\text{int}}$ and the
	inter-estimator ratios are independent of it over the whole plausible range
	$\phi_{\text{int}} \in [0.05,\,0.50]$.
	
	\subsection{Uncertainty budget}
	\label{sec:uncertainty}
	
	Table~\ref{tab:uq} collects the uncertainty contributions to
	$k_{\text{surf}} \cdot K_{\text{ads}}$ at $\phi = 0.70$. The purely numerical
	terms are negligible: the MC statistical standard error at the production photon
	budget is 0.005\% and the seed-to-seed morphological variability, from three
	independent Cahn--Hilliard realisations, is 0.03\% (Section~S4 of the
	Supplementary Information). The budget is dominated by two systematic terms: the
	choice of effective-medium rule, which shifts the descriptor by 4.3\% between
	Maxwell--Garnett and Bruggeman (Section~S5), and the experimental uncertainty on
	$k_{\text{app}}$, taken as 5\% to cover the pseudo-first-order fit standard error
	and sample reproducibility. A $\pm 20$\,\% bracket on $\mu_s'$, introduced in
	Section~\ref{sec:optics} to cover neglected dependent-scattering corrections,
	shifts the absolute descriptor by at most 1.2\% (Section~S5) and is included as
	a systematic term. This term, like the mixing-rule term, affects the
	\emph{absolute} extraction; the inter-estimator ratios that carry the
	morphological conclusions move by less than 7\% under the same perturbations
	(Table~S5 of the Supplementary Information) and are therefore substantially more
	robust than the absolute value of $k_{\text{surf}}\cdot K_{\text{ads}}$.
	
	\begin{table}[htbp]
		\centering
		\caption{Uncertainty budget for the \emph{absolute} value of
			$k_{\text{surf}} \cdot K_{\text{ads}}$
			(MC~3D estimator, $\phi = 0.70$). Terms are combined in quadrature.
			Inter-estimator ratios are substantially more robust: they move by less than
			7\% under the same optical perturbations (Table~S5 of the Supplementary
			Information).}
		\label{tab:uq}
		\begin{tabular}{lc}
			\toprule
			Source & Relative contribution \\
			\midrule
			Monte-Carlo statistics ($N = 1.85\times10^8$) & 0.005\% \\
			Morphological realisation (3 Cahn--Hilliard seeds) & 0.03\% \\
			Effective-medium rule (MG $\leftrightarrow$ BR, systematic) & 4.3\% \\
			Dependent scattering ($\mu_s'\ \pm 20$\%, systematic) & 1.2\% \\
			Experimental $k_{\text{app}}$ (fit and reproducibility) & 5.0\% \\
			\midrule
			Combined & \textbf{6.7\%} \\
			\bottomrule
		\end{tabular}
	\end{table}
	
	The extracted value is therefore
	\begin{equation}
		k_{\text{surf}} \cdot K_{\text{ads}} = (3.51 \pm 0.24) \times 10^{-35}
		\ \text{m}^4\,\text{photon}^{-1}
		\qquad (\text{MC~3D},\ \phi = 0.70).
		\label{eq:ksKads_final}
	\end{equation}
	This 6.7\% uncertainty should be compared with the factor of 1.9 separating the
	three estimators: it is the choice of optical model, and not numerical precision
	or experimental scatter, that limits the reliability of kinetic descriptors
	extracted from bulk photocatalytic measurements.
	
	\section{Discussion}
	\label{sec:discussion}
	
	\subsection{Physical mechanism: optical phase contrast}
	\label{sec:phase_contrast}
	
	The results of Section~\ref{sec:mechanism} identify the origin of the
	morphological concentration effect unambiguously. The two phases differ not only
	in absorption but, more importantly here, in scattering: the pores carry
	$\mu_a = \mu_s = 0$ while the solid carries $\mu_s' = 20118$~m$^{-1}$. A photon
	entering a void crosses it along a straight line and leaves it quickly, whereas
	in the solid it performs a random walk and accumulates a much longer path per
	unit volume. Since the fluence is precisely the path density per unit volume, the
	scattering phase sustains the higher local fluence. The measured contrast,
	$r = \langle\Phi\rangle_{\text{pore}}/\langle\Phi\rangle_{\text{solid}} = 0.514$,
	is the direct expression of this asymmetry.
	
	Two consequences follow. First, the catalytic phase is the bright phase: the
	conventional volume average, which treats all voxels alike, systematically
	underestimates the photon dose actually experienced by the TiO$_2$ nanoparticles,
	by 51.5\% at $\phi = 0.70$ and 77.4\% at $\phi = 0.90$. Second, the
	porosity dependence of that bias appears compositional rather than morphological
	for the spinodal geometries investigated. The contrast $r$ is the same at both
	simulated porosities to within 0.2\%, so Eq.~\eqref{eq:dilution} accounts for the
	whole trend: raising $\phi$ simply increases the statistical weight of the dark
	phase in the volume average. The phase-average identity is exact for each geometry. The near-equality of $r$
	at the two investigated porosities is an empirical observation under the present
	optical and morphological conditions. Whether a common value of $r$ can be used
	for other porosities, pore-size distributions, or morphology classes requires
	further tests.
	
	\subsection{Role of the void network in transport}
	\label{sec:channelling}
	
	The void network nevertheless plays a definite role, but in \emph{transport}
	rather than in local fluence. Because pores neither absorb nor scatter, they
	provide low-attenuation shortcuts through the medium, and the path statistics of
	Fig.~\ref{fig:channelling}c quantify how effective these shortcuts are: the mean
	cumulative in-pore path per escaping photon is 93.8~$\mu$m at $\phi = 0.70$ and
	276.5~$\mu$m at $\phi = 0.90$, and a weight fraction of 0.139 at the higher
	porosity exceeds the slab thickness itself. This is a genuinely
	multiply-connected, percolating void phenomenon, geometrically underpinned by the
	long tail of the chord distribution (Fig.~\ref{fig:channelling}a) and enhanced by
	the secondary population of the PSD (Fig.~\ref{fig:morphology}b).
	
	The distinction between the two effects matters. Transport through the void
	network makes the depth profile non-diffusive and is the principal reason why the
	DA fails at the profile level (Section~\ref{sec:homogeneous}); the phase contrast
	makes the volume average a biased estimator of the catalytic photon dose. They
	act on different observables and must not be conflated. In particular, the local
	fluence inside a pore does not increase with pore size---the co-localisation
	correlation is $\rho = +0.03$, i.e.\ negligible---so no design rule should be
	built on the notion of bright large voids.
	
	\subsection{Homogeneous-medium simulations: isolating the morphological contribution}
	\label{sec:homogeneous}
	
	To separate the intrinsic error of the DA from the additional error introduced by
	the discrete morphology, MC simulations were performed on two homogeneous
	(non-spinodal) slabs of identical thickness $L = 0.557$~mm, with the same
	boundary treatment. Full profiles are shown in Figs.~S2 and S3 of the
	Supplementary Information and the metrics are collected in Table~S1.
	
	\paragraph{Homogeneous solid slab} The domain is filled with the shell
	coefficients ($\mu_a = 5159$~m$^{-1}$, $\mu_s' = 20118$~m$^{-1}$,
	$\kappa L = 11.0$, $L/\ell^* = 11.2$), placing the system well inside the
	diffusion regime. The collimated DA reproduces the MC profile with
	$\mathrm{RMS}_w = 6.3$\%, confirming that the boundary treatment and the DA
	formulation are internally consistent. The residual deviation arises from the
	moderate absorption-to-scattering ratio, not from the boundary conditions.
	
	\paragraph{Homogeneous EMA slab} The domain is filled with the EMA coefficients
	($\mu_a = 1548$~m$^{-1}$, $\mu_s' = 6035$~m$^{-1}$, $\kappa L = 3.30$). This
	system has exactly the optical thickness of the spinodal benchmark but none of
	its discrete structure. The DA now deviates by $\mathrm{RMS}_w = 8.5$\%, the
	degradation relative to the solid slab being attributable to the lower optical
	thickness, which moves the system further from the diffusion regime.
	
	\paragraph{Decomposition} The matched-$\kappa L$ comparison is the relevant one:
	$\mathrm{RMS}_w=8.5\%$ for the homogeneous EMA slab and $54.6\%$ for the
	spinodal geometry. The difference between these two empirical metrics indicates
	that the homogeneous DA benchmark does not adequately represent photon transport
	in the simulated heterogeneous structure. It is not an additive error
	decomposition, and it does not imply that no other choice of effective
	coefficients could reduce the homogeneous-slab discrepancy. The 8.5\% residual
	at these optical parameters still propagates into depth-averaged quantities and
	should be borne in mind when the DA is used as an engineering approximation.
	
	\subsection{Comparison with experimental optical measurements}
	\label{sec:experimental_optics}
	
	The corrected photon budget makes falsifiable optical predictions---diffuse
	reflectance $R = 0.354$ and transmittance $T = 0.003$ at $\phi = 0.70$
	($R = 0.518$, $T = 0.052$ at $\phi = 0.90$; Table~\ref{tab:energy}), together
	with the decay profiles of Fig.~\ref{fig:fluence_profile}. These can be
	confronted with the measurements already published for the companion
	TiO$_2$-loaded monolith series~\cite{Layan2025}. A consolidated
	prediction-versus-measurement table, covering both the morphological targets of
	Section~\ref{sec:morphology} and the optical quantities discussed here, is given
	in Section~S6 of the Supplementary Information.
	
	\paragraph{Diffuse reflectance at 365~nm} Read at the operating wavelength,
	which lies on the steep absorption edge, the UV--vis diffuse-reflectance spectra
	of Ref.~\cite{Layan2025} give $R \approx 0.54$ for a Ti/Si molar ratio
	$x = 0.25$ and $R \approx 0.48$ for $x = 0.40$, rising to $R \approx 0.8$ for the
	least loaded member, with deep-UV plateau minima of $\approx 0.19$--$0.20$ for
	the loaded members; the wavelength registration was cross-checked against the
	365-nm lamp emission line displayed in the same figure. At the porosity
	representative of the real Si(HIPE) monoliths ($\phi \approx 0.9$), the predicted
	front-face reflectance $R = 0.518$ therefore agrees with the 365-nm measurements on
	the most loaded members to within a few percent, while the denser benchmark
	($\phi = 0.70$, $R = 0.354$) lies below them, as expected from its larger solid
	fraction and higher absorbance. Monte Carlo $R$ is the total front-face reflectance (entry Fresnel plus later
	escape), not strictly diffuse reflectance. At $\phi = 0.90$ the estimated
	specular share is $\lesssim 1$\%, so the comparison with the UV--vis values
	$R \approx 0.48$--$0.54$ remains a fair consistency check but is not labelled as
	diffuse-to-diffuse without comment.
	
	\paragraph{Sources of residual difference} Three effects separate the model from
	the measurement, all of them acting in the observed direction.
	(i)~\emph{Composition mapping}: the molar ratios $x = 0.25$--$0.40$ correspond to
	NP volume fractions $f_{\text{NP}} \approx 0.16$--$0.23$, slightly below the
	simulated $f_{\text{NP}} = 0.25$, so that the real samples absorb somewhat less
	and reflect somewhat more. (ii)~\emph{Hierarchical porosity}: the walls of the
	real monoliths carry a micro/mesoporous texture (BET
	700--1200~m$^2$\,g$^{-1}$~\cite{Layan2025}) which the two-phase model does not
	represent and which adds scattering interfaces inside the solid.
	(iii)~\emph{Dependent scattering}: the independent-scatterer estimate of $\mu_s'$
	(Section~\ref{sec:optics}) neglects correlations that are not negligible at
	$f_{\text{NP}} = 0.25$.
	
	\paragraph{Penetration depth} The Au$^0$-photoreduction experiment of
	Ref.~\cite{Layan2025} visualises the UV penetration decreasing from $>1$~cm to
	$\approx 6$~mm as the TiO$_2$ loading increases. This is the direct signature of
	the loading and porosity dependence of $\kappa_{\text{eff}}$ obtained here, and
	of the deeper penetration predicted at higher porosity, where the front/back
	fluence ratio falls from 318 to 25 between $\phi = 0.70$ and $\phi = 0.90$.
	
	\paragraph{Status of the comparison} Three levels of evidence should be
	distinguished.
	(i)~\emph{Numerical conservation and internal consistency} are validated per run
	by the independently tallied closure $R+T+A=1$ to better than $10^{-7}$
	(Section~\ref{sec:MC}), by the homogeneous-slab DA tests of
	Section~\ref{sec:homogeneous}, and by the convergence studies of Section~S4 of
	the Supplementary Information.
	(ii)~\emph{Macroscopic reflectance} at $\phi \approx 0.9$ agrees with previously
	published 365-nm UV--vis measurements on a related monolith series to within a
	few percent. The two simulated porosities bracket rather than reproduce the
	experimental compositions, $f_{\text{NP}}$ is not matched, and Monte Carlo $R$
	is total front-face reflectance rather than strictly diffuse reflectance. We
	treat this as a consistency check on the photon budget, not as a
	parameter-matched calibration.
	(iii)~\emph{Direct experimental validation of the predicted 3D internal fluence
	field}---and therefore of the extracted kinetic descriptor---is not claimed:
	the geometries are Cahn--Hilliard realisations rather than specimen-specific
	reconstructions, and no spatially resolved internal fluence measurement is
	available.
	The estimator ratios that carry the conclusions of this work are, however,
	insensitive to the absolute fluence level (Section~S5 of the Supplementary
	Information).
	
	\subsection{Critical role of effective-medium consistency}
	\label{sec:EMA_consistency}
	
	The requirement to scale both $\mu_a$ and $\mu_s'$ by $(1-\phi)$
	(Eq.~\eqref{eq:EMA_scaling}) is straightforward but easily overlooked. Using the
	shell $\mu_s' = 20118$~m$^{-1}$ together with $\mu_{a,\text{eff}} =
	1548$~m$^{-1}$ yields
	\begin{equation}
		\frac{\kappa_{\text{wrong}}}{\kappa_{\text{correct}}}
		= \sqrt{\frac{\mu_{a,\text{eff}} + \mu_{s,\text{shell}}'}
			{\mu_{a,\text{eff}} + \mu_{s,\text{eff}}'}}
		= \sqrt{\frac{21666}{7583}} \approx 1.69,
		\label{eq:kappa_wrong}
	\end{equation}
	overestimating the decay constant by 69\%, compressing the predicted profile and
	reducing $z_b$ from 110~$\mu$m to 33~$\mu$m---the shell value, which is
	physically meaningless at the domain level. This consistency condition must be
	verified explicitly whenever optical coefficients from shell-level calculations
	or from the literature are used to construct DA parameters for a porous medium.
	
	\subsection{Validity and limits of the diffusion approximation}
	\label{sec:DA_validity}
	
	The benchmark system ($\mu_a/\mu_s' = 0.256$, $\kappa_{\text{eff}} L = 3.30$)
	sits at the boundary of DA validity on two counts. First,
	$\mu_a/\mu_s' = 0.256$ violates $\mu_a \ll \mu_s'$ ($< 0.05$
	recommended~\cite{Ishimaru1978}), so the DA overestimates diffuse contributions
	and misses quasi-ballistic transport. Second, $\kappa L = 3.30$ is only
	marginally diffusive ($\kappa L > 5$ typically required~\cite{Haskell1994}). The homogeneous-medium comparisons of Section~\ref{sec:homogeneous} show that
	these two factors produce an $\mathrm{RMS}_w$ of 8.5\% for a homogeneous EMA
	slab of the same optical thickness, against 54.6\% for the spinodal geometry.
	As stated there, this comparison of empirical metrics is not an additive
	decomposition.
	
	At the level of the depth-averaged fluence, the DA overestimates
	$\langle\Phi\rangle$ by 25\% relative to the MC~3D reference (MC~3D/DA $= 1.254$
	on the extracted descriptor, i.e.\ a 20\% underestimate of
	$k_{\text{surf}} \cdot K_{\text{ads}}$). For engineering applications where
	$\sim 25$\% accuracy is acceptable, the DA with full EMA consistency therefore
	remains a useful screening tool. For accurate extraction of
	$k_{\text{surf}} \cdot K_{\text{ads}}$, or for optimisation of the NP
	distribution, the MC~3D solid-weighted estimator is indispensable.
	
	\subsection{Effect of porosity}
	\label{sec:porosity}
	
	The Supplementary Information presents the complete $\phi = 0.90$ case (Figs.~S4,
	S5 and S6, Table~S2), whose headline quantities are also carried in
	Tables~\ref{tab:energy} and~\ref{tab:kinetics} of the main text. Raising the
	porosity has two coupled consequences. The EMA coefficients scale as $(1-\phi)$
	and the optical thickness falls to $\kappa_{\text{eff}} L = 1.10$, placing the
	sample outside the diffusion regime; and the solid skeleton becomes thinner and
	more fragmented, with the mean solid chord dropping from 8.1 to 2.6~$\mu$m while
	the mean pore chord grows from 18.6 to 22.8~$\mu$m.
	
	The concentration ratio rises from 1.515 to 1.774, entirely as predicted by
	Eq.~\eqref{eq:dilution} at constant $r$. The MC~3D/DA ratio, in contrast, falls
	from 1.254 to 1.022. This apparent improvement of the DA is misleading: at
	$\kappa L = 1.10$ the DA is further from validity than at 3.30, and its
	profile-level discrepancy remains large ($\mathrm{RMS}_w = 51.1$\%). What happens
	is that the depth-averaged error becomes small by cancellation, because the
	overall attenuation is weak. Neither porosity gives DA predictions that can be
	trusted at the profile level, and the near-unit ratio at $\phi = 0.90$ should not
	be read as a validation.
	
	\subsection{Implications for photocatalytic reactor design}
	\label{sec:implications}
	
	Three practical implications follow.
	
	\paragraph{Solid fraction optimisation} Increasing porosity amplifies the
	concentration ratio, from 1.515 at $\phi = 0.70$ to 1.774 at $\phi = 0.90$, and
	therefore the photon dose per catalytic site. It simultaneously reduces the
	accessible catalytic surface, which scales as $(1-\phi)$, and the absorptance,
	which falls from 0.643 to 0.430. Equation~\eqref{eq:dilution} now makes the first
	of these three effects analytically predictable, so that the optimum can be
	sought without a full 3D simulation at every porosity---although the transport
	terms still require one.
	
	\paragraph{Pore-size distribution engineering} The Cahn--Hilliard stopping
	criterion controls the extent of coarsening and hence the weight of the
	secondary PSD population. A broader distribution lengthens the tail of the
	pore-chord distribution and, in the two geometries simulated here, accompanies
	more low-attenuation void transport, which extends the irradiated depth. Whether
	that change leaves the phase-contrast ratio $r$ unaltered, and therefore whether
	penetration depth and catalytic photon dose can be tuned independently at fixed
	porosity, is not established by the present runs and would require a dedicated
	parametric study. This is therefore a design hypothesis, not a demonstrated
	rule.
	
	\paragraph{NP distribution mode} The present results assume bulk-distributed
	NPs. Because the fluence is highest inside the scattering solid rather than at
	the pore surface, localising the NPs at the solid/pore interface would place them
	in a region of \emph{lower} local fluence than the ligament interior, while
	improving their accessibility to the pollutant, i.e.\ raising
	$\chi_{\text{access}}$ in Eq.~\eqref{eq:ksurf_micro}. The optimum therefore
	involves a genuine trade-off between photon dose and reactant access, which the
	present pipeline is able to quantify and which we leave to future work.
	
	\section{Conclusions}
	\label{sec:conclusions}
	
	We have presented a complete computational pipeline---Cahn--Hilliard spinodal
	geometry generation, Maxwell--Garnett effective-medium optical properties, 3D GPU
	Monte Carlo transport, and a collimated-source diffusion-approximation
	benchmark---applied to bicontinuous spinodal aerogel photocatalysts at
	$\lambda = 365$~nm. Our conclusions are of two kinds.
	
	\textbf{Technically}, we have shown that (1)~specular mirrors on the lateral
	faces with open axial faces and a one-voxel air pad, combined with independent
	tallies of $R$, $T$ and $A$, produce a photon budget that closes to better than
	$10^{-7}$ (Eq.~\eqref{eq:closure}); because no channel is obtained by difference,
	this closure is a genuine per-run validation and it is satisfied by all
	simulations reported here. (2)~Both $\mu_a$ and $\mu_s'$ must be scaled by
	$(1-\phi)$ when constructing DA parameters; using the shell $\mu_s'$ with the EMA
	$\mu_a$ overestimates $\kappa$ by 69\% and compresses the fluence profile.
	(3)~The correct DA for planar illumination decomposes the fluence into ballistic
	and diffuse terms (Eqs.~\eqref{eq:DA_total}--\eqref{eq:Phi_diff}); the
	isotropic-source Green's function~\cite{Vallee2026} introduces a spurious peak at
	$z = \ell^*$ and should not be used for collimated sources. (4)~The unweighted
	relative RMS is a tail-dominated and potentially misleading measure of MC--DA
	agreement; the fluence-weighted variant of Eq.~\eqref{eq:RMSw} should be
	preferred.
	
	\textbf{Scientifically}, we have found that (1)~the energy balance gives
	$R = 0.354$, $T = 0.003$, $A = 0.643$ at $\phi = 0.70$ and $R = 0.518$,
	$T = 0.052$, $A = 0.430$ at $\phi = 0.90$. (2)~The catalytic solid phase
	experiences 51.5\% more photons than the volume average at $\phi = 0.70$, rising
	to 77.4\% at $\phi = 0.90$. (3)~The mechanism is an optical phase contrast: with
	$\mu_a = \mu_s = 0$ in the voids, photons cross them quasi-ballistically and the
	pore phase sustains only $r = 0.514$ of the solid-phase fluence. The ratio $r$ is
	unchanged to within 0.2\% between the two simulated porosities---an empirical
	result for the conditions investigated. The concentration ratio then follows
	from the dilution identity of Eq.~\eqref{eq:dilution}; reproducing it from the
	measured $r$ is algebraic consistency, not an independent prediction.
	Constancy of $r$ under changes of porosity, pore-size distribution, morphology
	class or optical coefficients is not established. (4)~The three kinetic estimators
	give $k_{\text{surf}} \cdot K_{\text{ads}}$ values spanning a factor of 1.9; the
	MC~3D solid-weighted value, $(3.51 \pm 0.24) \times 10^{-35}$~m$^4$\,photon$^{-1}$,
	is the highest-fidelity reference considered in this study, the DA underestimating it by 20\% and the volume
	average overestimating it by 51.5\%. (5)~Matched-$\kappa L$ homogeneous-slab comparisons give
	$\mathrm{RMS}_w = 8.5\%$ for a homogeneous EMA slab and $54.6\%$ for the
	spinodal geometry. The difference between these two empirical metrics indicates
	that the homogeneous DA benchmark does not adequately represent photon transport
	in the simulated heterogeneous structure. It is not an additive error
	decomposition, and it does not imply that no other choice of effective
	coefficients could reduce the homogeneous-slab discrepancy. (6)~The void
	network provides low-attenuation transport paths: at $\phi = 0.90$, 14\% of the
	escaping photon weight has travelled further inside the void phase than the slab
	is thick. This affects penetration depth and profile shape, but not the local
	fluence contrast, and it cannot be captured by any 1D model. (7)~The predicted front-face reflectance at $\phi \approx 0.9$
	($R = 0.518$) agrees with the 365-nm UV--vis measurements on the most
	loaded members of the companion monolith series ($R \approx 0.48$--$0.54$)
	to within a few percent. Following Section~\ref{sec:experimental_optics},
	this is a consistency check on the macroscopic photon budget, not a
	validation of the internal 3D fluence field. The predicted porosity
	dependence of the penetration depth is consistent with the
	Au$^0$-photoreduction imaging of the same series.
	
	Together with the design implications of Section~\ref{sec:implications}, these
	results provide quantitative guidance for the use of highly porous aerogels as
	volumetric photocatalysts, and a correction that should be applied whenever
	intrinsic kinetic descriptors are extracted from bulk measurements on
	bicontinuous materials.
	
	\appendix
	
	\section{Cahn--Hilliard numerical parameters}
	\label{app:CH}
	
	\paragraph{Initial condition} $\varphi(\mathbf{x},0) = 0.5 + \eta(\mathbf{x})$
	with $\eta \sim \mathcal{N}(0,\,0.05^2)$ independently per voxel, clipped to
	$[0,1]$. The pseudo-random generator is seeded explicitly; the production
	geometries use seed~0 and the three-realisation ensemble of Table~S3 uses
	seeds~1--3. The integration is always performed at the spinodal centre
	$\varphi_{\text{sim}} = 0.5$, the target porosity being imposed only at
	binarisation.
	
	\paragraph{Gradient coefficient and time step} For
	$d_{\text{target}} = 17.4$~$\mu$m and $a_{\text{vox}} = 1.45$~$\mu$m,
	Eq.~\eqref{eq:kappa_cal} gives $\kappa_{\text{CH}} = 0.1140$ in voxel units. The
	semi-implicit scheme of Eq.~\eqref{eq:CH_spectral} is unconditionally stable with
	respect to the stiff interfacial term, so the step is set by the explicit
	nonlinear term:
	$\Delta t = \min[0.5/(\kappa_{\text{CH}}k_{\max}^4),\,1] = 0.0450$ in
	dimensionless Cahn--Hilliard units, with $k_{\max} = \pi$~voxel$^{-1}$.
	
	\paragraph{Stopping criterion} The dominant structural size is measured every
	200 steps as the peak of the radial power spectral density; integration stops
	when it first enters
	$[d_{\text{target}} - \sigma/2,\ d_{\text{target}} + \sigma/2] =
	[14.8,\,20.0]$~$\mu$m, with a hard cap of 50\,000 steps.
	
	\paragraph{Binarisation} The binary mask is obtained by rank-based thresholding
	of the final field: the $\phi\,N_{\text{vox}}$ voxels of highest $\varphi$ are
	labelled pore, with explicit tie-breaking. This enforces the target porosity
	exactly and, unlike quantile thresholding, is robust to the ties that arise when
	a coarsened field is block-averaged. The achieved porosities are $0.7000$ and
	$0.9000$.
	
	\section{Monte Carlo solver configuration}
	\label{app:MC}
	
	\paragraph{Volume and media} The transport volume is the binary mask with pore
	voxels labelled medium~1 ($\mu_a = \mu_s = 0$, $n = 1$) and solid voxels
	medium~2 (shell coefficients of Table~\ref{tab:optics},
	$n_{\text{eff}} = 1.746$), padded by one voxel of air (medium~0) on each axial
	face. Voxel edge $a_{\text{vox}} = 1.45$~$\mu$m.
	
	\paragraph{Source} Planar collimated source covering the whole $z = 0$ face,
	propagating along $+z$.
	
	\paragraph{Boundary conditions} \texttt{bc = "mmamma"}: specular mirrors on
	$\pm x$ and $\pm y$, open on $\pm z$. Internal Fresnel refraction and reflection
	are enabled at all index-mismatched interfaces (\texttt{isreflect = 1}); no
	separate specular source term is used (\texttt{isspecular = 0}), the entry
	reflection being carried by the air pad.
	
	\paragraph{Tallies} Volumetric fluence and the boundary escape map
	(\texttt{issaveref = 1}) are both recorded. $A$ is taken from the absorbed-energy counter and $R$, $T$ from the
	escape map on the two air-pad planes, according to Eq.~\eqref{eq:RTA}. The
	closure of Eq.~\eqref{eq:closure} is verified for every run.
	
	\paragraph{Statistics} $N = 1.85 \times 10^{8}$ launched packets for all
	production runs; fixed random seed 20260809; weighted-photon transport with
	Russian-roulette termination. Time gate from Eq.~\eqref{eq:tend}. Convergence
	data are reported in Section~S4 of the Supplementary Information.
	
	\paragraph{Post-processing} Fluence normalisation follows
	Eq.~\eqref{eq:Phi_norm}. The three estimators are computed from
	Eqs.~\eqref{eq:Phi_MC3D}, \eqref{eq:Phi_MCvol} and~\eqref{eq:Lopt_DA}; the
	phase-resolved profiles of Fig.~\ref{fig:channelling}b are obtained by averaging
	$\Phi$ separately over solid and pore voxels at each depth and normalising by the
	lateral mean at the same depth.
	
	\section*{CRediT authorship contribution statement}
	
	\textbf{Renaud A.L. Vall\'ee:} Conceptualisation, Methodology, Software, Formal
	analysis, Investigation, Writing -- original draft, Writing -- review and
	editing, Visualisation.
	
	\section*{Declaration of competing interest}
	
	The author declares no competing financial interests.
	
	\section*{Data availability}
	
	A Supplementary Information file accompanies this article. The geometry masks and
	the master results file will be made available on reasonable request.
	
	\section*{Acknowledgements}
	
	The author acknowledges the computing resources provided at CRPP.


\begin{thebibliography}{99}
		
		\bibitem{Hoffmann1995}
		M.R.\ Hoffmann, S.T.\ Martin, W.\ Choi, D.W.\ Bahnemann,
		Environmental applications of semiconductor photocatalysis,
		Chem.\ Rev.\ 95 (1995) 69--96.
		\url{https://doi.org/10.1021/cr00033a004}
		
		\bibitem{Vardon2025}
		A.\ Vardon, M.-A.\ Dourges, \'E.\ Laurichesse, V.\ Schmitt, A.\ Bentaleb,
		F.\ Nallet, I.\ Ly, R.\ Backov,
		Mastering syntheses of siliceous hierarchical porous self-standing monoliths
		through the integration of the sol--gel process, complex fluids, and a planetary
		mixer,
		Langmuir 41 (2025) 26760--26769.
		\url{https://doi.org/10.1021/acs.langmuir.5c03237}
		
		\bibitem{Layan2025}
		\'E.\ Layan, G.\ Clermont, I.\ Ly, F.\ Nallet, R.A.L.\ Vall\'ee, et al.,
		Bulk photocatalysis within anatase-TiO$_2$--SiO$_2$ monoliths for efficient VOC
		photodegradation,
		Chem.\ Commun.\ 61 (2025) 19269--19272.
		\url{https://doi.org/10.1039/d5cc02552d}
		
		\bibitem{Ollis2018}
		D.F.\ Ollis,
		Kinetics of photocatalyzed reactions: Five lessons learned,
		Front.\ Chem.\ 6 (2018) 378.
		\url{https://doi.org/10.3389/fchem.2018.00378}
		
		\bibitem{Bloh2019}
		J.Z.\ Bloh,
		A holistic approach to model the kinetics of heterogeneous photocatalysis,
		Front.\ Chem.\ 7 (2019) 128.
		\url{https://doi.org/10.3389/fchem.2019.00128}
		
		\bibitem{Garnett1904}
		J.C.M.\ Garnett,
		Colours in metal glasses and in metallic films,
		Philos.\ Trans.\ R.\ Soc.\ A 203 (1904) 385--420.
		\url{https://doi.org/10.1098/rsta.1904.0024}
		
		\bibitem{Bruggeman1935}
		D.A.G.\ Bruggeman,
		Berechnung verschiedener physikalischer Konstanten von heterogenen Substanzen,
		Ann.\ Phys.\ 416 (1935) 636--664.
		\url{https://doi.org/10.1002/andp.19354160705}
		
		\bibitem{Ishimaru1978}
		A.\ Ishimaru,
		Wave Propagation and Scattering in Random Media,
		IEEE Press, New York, 1978.
		
		\bibitem{Vallee2026}
		R.A.L.\ Vall\'ee, R.\ Backov,
		From light diffusion to photocatalytic rates: Compact scaling laws for strongly
		scattering porous slabs,
		J.\ Quant.\ Spectrosc.\ Radiat.\ Transfer 352 (2026) 109819.
		\url{https://doi.org/10.1016/j.jqsrt.2026.109819}
		
		\bibitem{Vynck2023}
		K.\ Vynck, R.\ Pierrat, R.\ Carminati, L.S.\ Froufe-P\'erez, F.\ Scheffold,
		R.\ Sapienza, S.\ Vignolini, J.J.\ Sa\'enz,
		Light in correlated disordered media,
		Rev.\ Mod.\ Phys.\ 95 (2023) 045003.
		\url{https://doi.org/10.1103/RevModPhys.95.045003}
		
		\bibitem{Leyre2020}
		S.\ Leyre, E.\ Coutino-Gonzalez, J.J.\ Joos, J.\ Ryckaert, Y.\ Meuret,
		D.\ Poelman, P.F.\ Smet, G.\ Durinck, J.\ Hofkens, G.\ Deconinck, P.\ Hanselaer,
		Absolute determination of photoluminescence quantum efficiency using an
		integrating sphere setup,
		Rev.\ Sci.\ Instrum.\ 91 (2020) 013108.
		\url{https://doi.org/10.1063/1.5136060}
		
		\bibitem{Leyre2022}
		S.\ Leyre, B.\ Meuris, J.J.\ Joos, P.F.\ Smet, G.\ Deconinck, P.\ Hanselaer,
		Determination of the optical properties of turbid media using a single
		integrating sphere setup,
		Appl.\ Opt.\ 61 (2022) 792--802.
		\url{https://doi.org/10.1364/AO.445979}
		
		\bibitem{Brunser2023}
		S.S.\ Brunser, F.L.\ Bargardi, R.\ Libanori, N.\ Kaufmann, H.\ Braun,
		A.\ Steinfeld, A.R.\ Studart,
		Solar-driven redox splitting of CO$_2$ using 3D-printed hierarchically channelled
		ceria structures,
		Adv.\ Mater.\ Interfaces 10 (2023) 2300452.
		\url{https://doi.org/10.1002/admi.202300452}
		
		\bibitem{Zhang2026}
		B.W.\ Zhang, J.B.\ Pu, T.\ Hu, et al.,
		AI-powered nonlinear optical imaging reveals protein spatial homogenization as an
		indicator of impaired bone quality in type 2 diabetes,
		Opto-Electron.\ Adv.\ 9 (2026) 250312.
		\url{https://doi.org/10.29026/oea.2026.250312}
		
		\bibitem{Zhang2026PBM}
		B.\ Zhang, Z.\ Chen, W.\ Li, et al.,
		Frequency-specific transcranial photobiomodulation elicits complementary
		glial mechanisms for neurovascular protection and amyloid clearance in
		Alzheimer disease,
		Cyborg Bionic Syst.\ 7 (2026) 0551.
		\url{https://doi.org/10.34133/cbsystems.0551}

		\bibitem{Liu2022}
		Q.\ Liu, K.\ Pan, Y.\ Lu, X.\ Zhang,
		Data-driven for accelerated design strategy of photocatalytic degradation
		activity prediction of doped TiO$_2$ photocatalyst,
		J.\ Water Process Eng.\ 49 (2022) 103126.
		\url{https://doi.org/10.1016/j.jwpe.2022.103126}
		
		\bibitem{Cahn1958}
		J.W.\ Cahn, J.E.\ Hilliard,
		Free energy of a nonuniform system. I. Interfacial free energy,
		J.\ Chem.\ Phys.\ 28 (1958) 258--267.
		\url{https://doi.org/10.1063/1.1744102}
		
		\bibitem{Cahn1959}
		J.W.\ Cahn,
		Free energy of a nonuniform system. II. Thermodynamic basis,
		J.\ Chem.\ Phys.\ 30 (1959) 1121--1124.
		\url{https://doi.org/10.1063/1.1730145}
		
		\bibitem{Cahn1961}
		J.W.\ Cahn,
		On spinodal decomposition,
		Acta Metall.\ 9 (1961) 795--801.
		\url{https://doi.org/10.1016/0001-6160(61)90182-1}
		
		\bibitem{Provatas2010}
		N.\ Provatas, K.\ Elder,
		Phase-Field Methods in Materials Science and Engineering,
		Wiley-VCH, Weinheim, 2010.
		\url{https://doi.org/10.1002/9783527631520}
		
		\bibitem{Geslin2015}
		P.-A.\ Geslin, I.\ McCue, B.\ Gaskey, J.\ Erlebacher, A.\ Karma,
		Topology-generating interfacial pattern formation during liquid metal dealloying,
		Nat.\ Commun.\ 6 (2015) 8887.
		\url{https://doi.org/10.1038/ncomms9887}
		
		\bibitem{Zhu1999}
		J.\ Zhu, L.-Q.\ Chen, J.\ Shen, V.\ Tikare,
		Coarsening kinetics from a variable-mobility Cahn--Hilliard equation: Application
		of a semi-implicit Fourier spectral method,
		Phys.\ Rev.\ E 60 (1999) 3564--3572.
		\url{https://doi.org/10.1103/PhysRevE.60.3564}
		
		\bibitem{Lifshitz1961}
		I.M.\ Lifshitz, V.V.\ Slyozov,
		The kinetics of precipitation from supersaturated solid solutions,
		J.\ Phys.\ Chem.\ Solids 19 (1961) 35--50.
		\url{https://doi.org/10.1016/0022-3697(61)90054-3}
		
		\bibitem{Wagner1961}
		C.\ Wagner,
		Theorie der Alterung von Niederschl\"agen durch Uml\"osen (Ostwald-Reifung),
		Z.\ Elektrochem.\ 65 (1961) 581--591.
		
		\bibitem{Okuta2017}
		R.\ Okuta, Y.\ Unno, D.\ Nishino, S.\ Hido, C.\ Loomis,
		CuPy: A NumPy-compatible library for NVIDIA GPU calculations,
		Proc.\ Workshop on Machine Learning Systems (LearningSys) at NeurIPS, 2017.
		
		\bibitem{Fang2009}
		Q.\ Fang, D.A.\ Boas,
		Monte Carlo simulation of photon migration in 3D turbid media accelerated by
		graphics processing units,
		Opt.\ Express 17 (2009) 20178--20190.
		\url{https://doi.org/10.1364/OE.17.020178}
		
		\bibitem{Boas2002}
		D.A.\ Boas, J.P.\ Culver, J.J.\ Stott, A.K.\ Dunn,
		Three dimensional Monte Carlo code for photon migration through complex
		heterogeneous media including the adult human head,
		Opt.\ Express 10 (2002) 159--170.
		\url{https://doi.org/10.1364/OE.10.000159}

		\bibitem{Yu2018}
		L.\ Yu, F.\ Nina-Paravecino, D.\ Kaeli, Q.\ Fang,
		Scalable and massively parallel Monte Carlo photon transport simulations for
		heterogeneous computing platforms,
		J.\ Biomed.\ Opt.\ 23 (2018) 010504.
		\url{https://doi.org/10.1117/1.JBO.23.1.010504}
		
		\bibitem{Haskell1994}
		R.C.\ Haskell, L.O.\ Svaasand, T.-T.\ Tsay, T.-C.\ Feng, M.S.\ McAdams,
		B.J.\ Tromberg,
		Boundary conditions for the diffusion equation in radiative transfer,
		J.\ Opt.\ Soc.\ Am.\ A 11 (1994) 2727--2741.
		\url{https://doi.org/10.1364/josaa.11.002727}
		
		\bibitem{AnataseOptical}
		A.\ Jolivet, C.\ Labbé, C.\ Frilay, O.\ Debieu, P.\ Marie, B.\ Horcholle,
		F.\ Lemarié, X.\ Portier, C.\ Grygiel, S.\ Duprey, W.\ Jadwisienczak,
		D.\ Ingram, M.\ Upadhyay, A.\ David, A.\ Fouchet, U.\ Lüders, J.\ Cardin,
		Structural, optical, and electrical properties of TiO$_2$ thin films
		deposited by ALD: Impact of the substrate, the deposited thickness and
		the deposition temperature,
		Appl.\ Surf.\ Sci.\ 608 (2023) 155214.
		\url{https://doi.org/10.1016/j.apsusc.2022.155214}
		
	\end{thebibliography}
\end{document}